\documentclass[preprints,article,accept,moreauthors,pdftex,10pt,a4paper]{mdpi}

\usepackage{booktabs}
\usepackage{multirow}
\usepackage{soul}
\usepackage{microtype}
\usepackage[labelformat=simple]{subfig}

\usepackage{upgreek}
\usepackage{units}
\usepackage{amssymb}
\usepackage{tabularx}
\usepackage{todonotes}
\graphicspath{{./Figures/pdf/}}
\def\ring#1{{\mathaccent'27 #1}}

\firstpage{1}
\pubvolume{10}
\doinum{}
\issuenum{9}
\articlenumber{0}
\pubyear{2018}
\copyrightyear{2018}
\history{Received: 1 September 2018; Accepted: 15 September 2018; Published: 21 September 2018}

\Title{{(Gravitational) Vacuum Cherenkov Radiation}}

\Author{Marco Schreck \orcidA{}}

\AuthorNames{Marco Schreck}

\address[1]{%
Departamento de F\'{\i}sica, Universidade Federal do Maranh\~{a}o (UFMA), Campus Universit\'{a}rio do Bacanga, S\~{a}o~Lu\'{\i}s~65080-805, Brazil; marco.schreck@gmx.de; Tel.: +55-(98)-3272-8293}

\abstract{This work reviews our current understanding of Cherenkov-type processes in vacuum that may occur due to a possible violation of Lorentz invariance. The description of Lorentz violation is based on the Standard Model Extension (SME). To get an overview as general as possible, the most important findings for vacuum Cherenkov radiation in Minkowski spacetime are discussed. After~doing so, special emphasis is put on gravitational Cherenkov radiation. For a better understanding, the essential properties of the gravitational SME are recalled in this context. The~common grounds and differences of vacuum Cherenkov radiation in Minkowski spacetime and in the gravity sector are emphasized.}


\begin{document}


\section{Introduction}

Cherenkov radiation emerges when a massive, electrically charged particle travels through an optical medium with a velocity $v$ that is larger than the phase velocity $c_{\mathrm{med}}$ of light in that medium. In fact, the characteristic, blueish glow that can be observed in the cooling water of nuclear reactors corresponds to this particular type of radiation. The effect was discovered by the Russians Cherenkov and Vavilov in 1934 \cite{Cherenkov:1934}. Just about three years later, a theoretical explanation of their observation was provided by Frank and Tamm in the context of classical electromagnetism \cite{Tamm:1937}. Cherenkov, Frank, and Tamm shared the Nobel prize of 1958. The fact that Vavilov was excluded is the likely explanation for why his name is usually omitted when referring to this phenomenon. The standard book reference on Cherenkov radiation is \cite{Jelley:1958}, which delivers explanations for various aspects of the effect. Although the book was first published several decades ago, most of the material covered is still valid and useful.

A microscopic explanation for Cherenkov radiation is as follows. As long as the charged particle travels with a speed $v<c_{\mathrm{med}}$, the few atoms or molecules polarized in the vicinity of the trajectory emit their wave trains out of phase leading to destructive interference. A particle propagating with a high velocity $v\geq c_{\mathrm{med}}$ enables the atoms or molecules of the medium to emit their wave trains in phase. Hence, wave trains interfere constructively producing coherent radiation far away from the source, which corresponds to the radiation observed. Constructive interference occurs on the surface of a Mach cone that bears the name Cherenkov cone. The radiation is emitted in directions {perpendicular} to the cone. These directions enclose an angle $\theta_c$ with the trajectory of the particle where $\theta_c$ is the Cherenkov angle. The latter is usually small, which means that Cherenkov radiation is emitted almost collinearly with the propagating particle. On the contrary, the Cherenkov cone is wide open.

Violations of Lorentz invariance are predicted by certain approaches {to} physics at the Planck scale~\cite{Kostelecky:1988zi,Gambini:1998it}. An effective description {of Lorentz violation}, which is supposed to be valid for energies much smaller than the Planck energy, is known as the Standard Model Extension (SME) \cite{Colladay:1996iz,Colladay:1998fq}. The SME parameterizes Lorentz and {\em CPT}-violation
in terms of background fields that are properly contracted with field operators. A background field remains fixed under Lorentz transformations of matter fields and it decomposes into controlling coefficients that describe the magnitude of Lorentz violation. The~existence of a Lorentz-violating background field for a photon can also be interpreted as a vacuum with a nontrivial refractive index. The latter can depend on photon energy, direction, and polarization leading to the effects of dispersion, anisotropy, and birefringence, respectively. It is expected that a nontrivial vacuum may be accompanied by the presence of some kind of Cherenkov-type effect. As~this effect occurs in vacuo, it is reasonable to refer to it as vacuum Cherenkov radiation.

In pre-relativity times, Sommerfeld was probably the first one to mention the concept of vacuum Cherenkov radiation \cite{Sommerfeld:1904a,Sommerfeld:1904b,Sommerfeld:1905}. Many decades later, Beall referred to it in the context of modified gravitational couplings \cite{Beall:1970rw}. {Coleman and Glashow presumably performed the first modern investigation in \cite{Coleman:1997xq}. The~purpose of the current article} is to provide a review on the research carried out on vacuum Cherenkov radiation since the point of time around the advent of the SME. For the past 20 years, a~large number of articles has appeared shedding light on particular aspects of this {interesting process \cite{Lehnert:2004be,Lehnert:2004hq,Kaufhold:2005vj,Altschul:2006zz,Altschul:2007kr,Kaufhold:2007qd,Klinkhamer:2008ky,Hohensee:2008xz,Cohen:2011hx,Altschul:2014bba,Schober:2015rya,Diaz:2015hxa,Kostelecky:2015dpa,Altschul:2016ces,Colladay:2016rmy,Schreck:2017isa,Colladay:2017qfr,DeCosta:2018nyf}. Studies have been carried out within classical field theory \cite{Lehnert:2004be,Lehnert:2004hq,Altschul:2006zz,Altschul:2007kr,Altschul:2014bba,Schober:2015rya,DeCosta:2018nyf} and quantum theory for spinless particles \cite{Kaufhold:2007qd} as well as elementary fermions \cite{Kaufhold:2005vj,Kaufhold:2007qd,Klinkhamer:2008ky,Hohensee:2008xz,Colladay:2016rmy,Schreck:2017isa}. The latter references deal with effective theories in the sense that the authors consider pointlike approximations of extended particles such as mesons, hadrons, and even atomic nuclei. The substructure of hadrons was taken into account in some papers where sophisticated analyses were carried out based on parton distribution functions of the proton and neutron \cite{Diaz:2015hxa,Colladay:2017qfr}.} It was also discovered that Cherenkov-type processes are not limited to the emission of photons {\cite{Cohen:2011hx,Altschul:2016ces,Colladay:2017qfr}}. For example, as neutrinos are not electrically charged, they cannot emit photons. However, they can still lose energy in the presence of Lorentz violation by radiating Z bosons {\cite{Cohen:2011hx}}. Last but not least, vacuum Cherenkov radiation can occur in the context of a modified gravity {\cite{Kostelecky:2015dpa}}. Lorentz violation in the pure-gravity sector modifies the dispersion relation and, therefore, the phase speed of gravitons. In a scenario where particles can travel faster than gravitons, they may lose energy by graviton emission. Special emphasis will be put on the SME gravity sector and the Cherenkov process associated with it.


\section{Fundamentals of the Process}
\label{sec:fundamentals-kinematics}

\begin{figure}[H]
\centering
\subfloat[]{\label{fig:vcr-kinematics-modified-maxwell-theory}\includegraphics[scale=0.2]{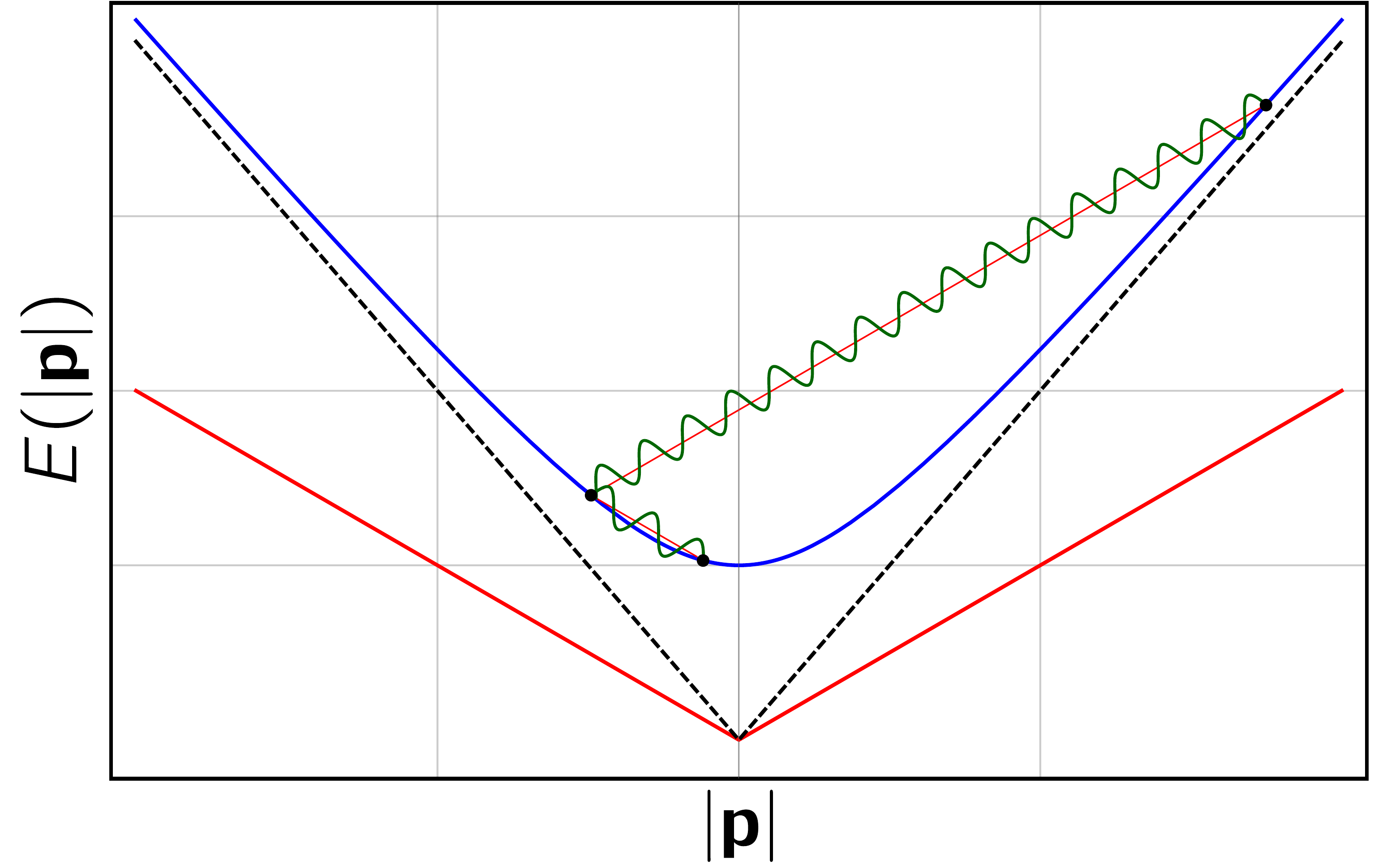}}\hspace{0.5cm}
\subfloat[]{\label{fig:vcr-kinematics-mcs-theory}\includegraphics[scale=0.2]{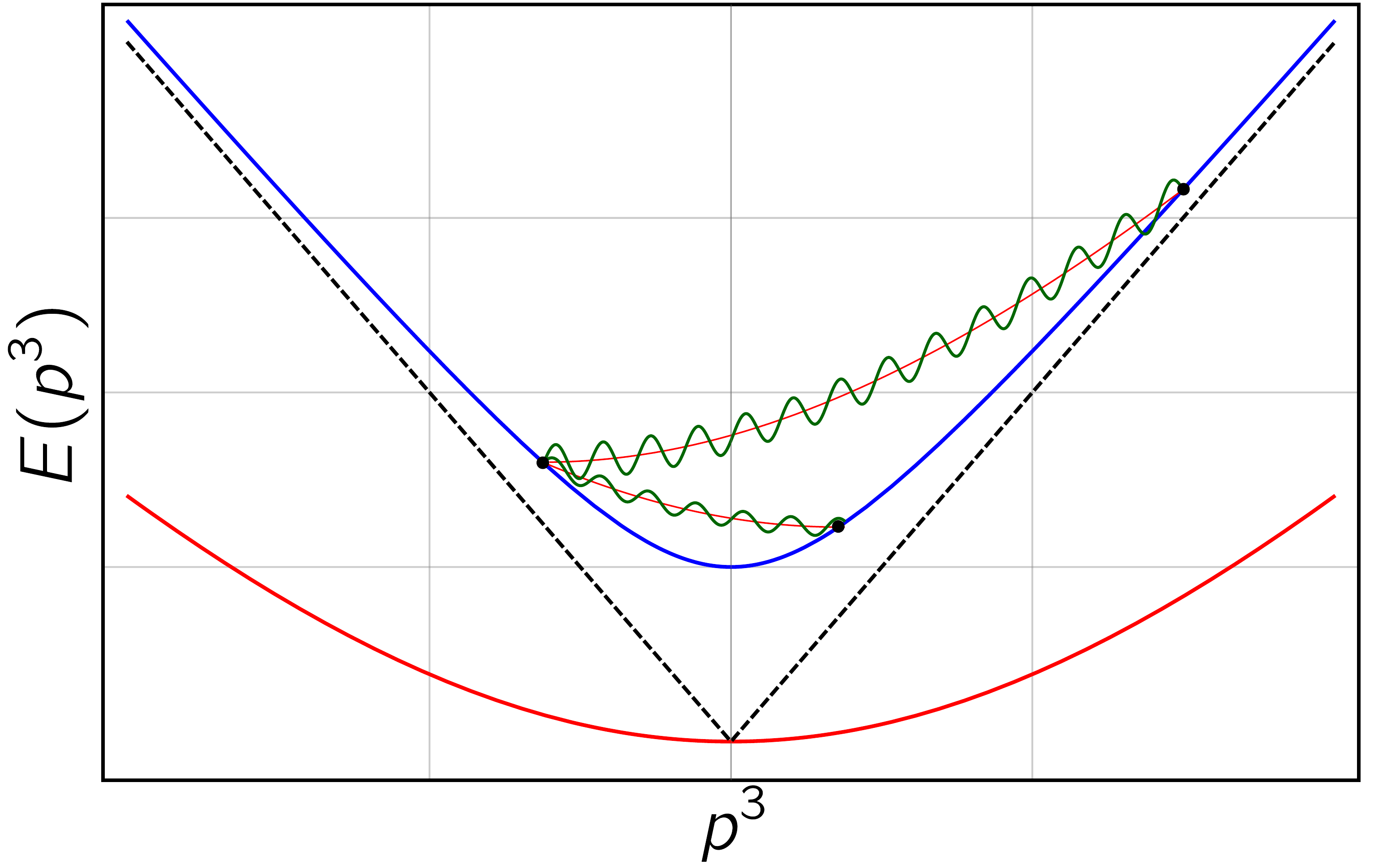}}
\caption{Standard nullcone (black, dashed), standard mass shell (blue), and modified photon dispersion relation (red) in momentum space with the momentum $\mathbf{p}=(p^1,p^2,p^3)$. Photon emissions from a standard Dirac fermion with a certain energy are indicated by green, wiggly curves. (\textbf{a}) {\em CPT}-even isotropic
modification of the photon sector with $\widetilde{\kappa}_{\mathrm{tr}}=3/5$. The wiggly lines run parallel to parts of the modified nullcone. In this particular example, two subsequent photon emissions are possible until the process ceases; (\textbf{b}) {\em CPT}-odd Maxwell--Chern--Simons (MCS) theory with $k_{AF}^3=5m_{\psi}$ {where} only one of the two modified photon dispersion laws {is} illustrated. The wiggly lines run along properly shifted parts of the deformed nullcone such that the minimum is at the lower one of the two points that it connects. Two subsequent photon emissions are shown.}
\label{fig:modified-photons}
\end{figure}

\begin{figure}[H]
\centering
\subfloat[]{\label{fig:vcr-kinematics-spin-degenerate-fermions}\includegraphics[scale=0.2]{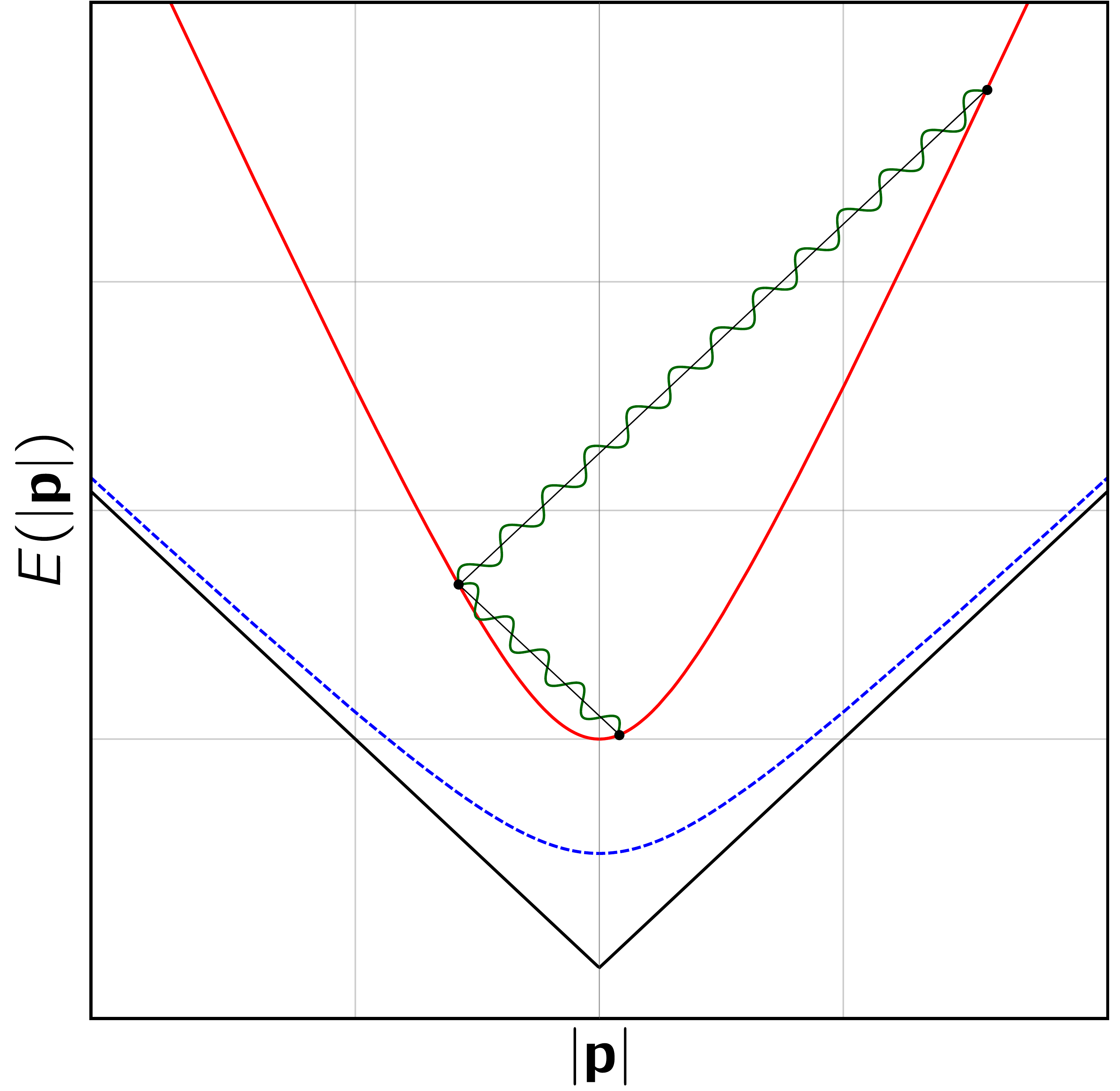}}\hspace{1cm}
\subfloat[]{\label{vcr-kinematics-spin-nondegenerate-fermions}\includegraphics[scale=0.2]{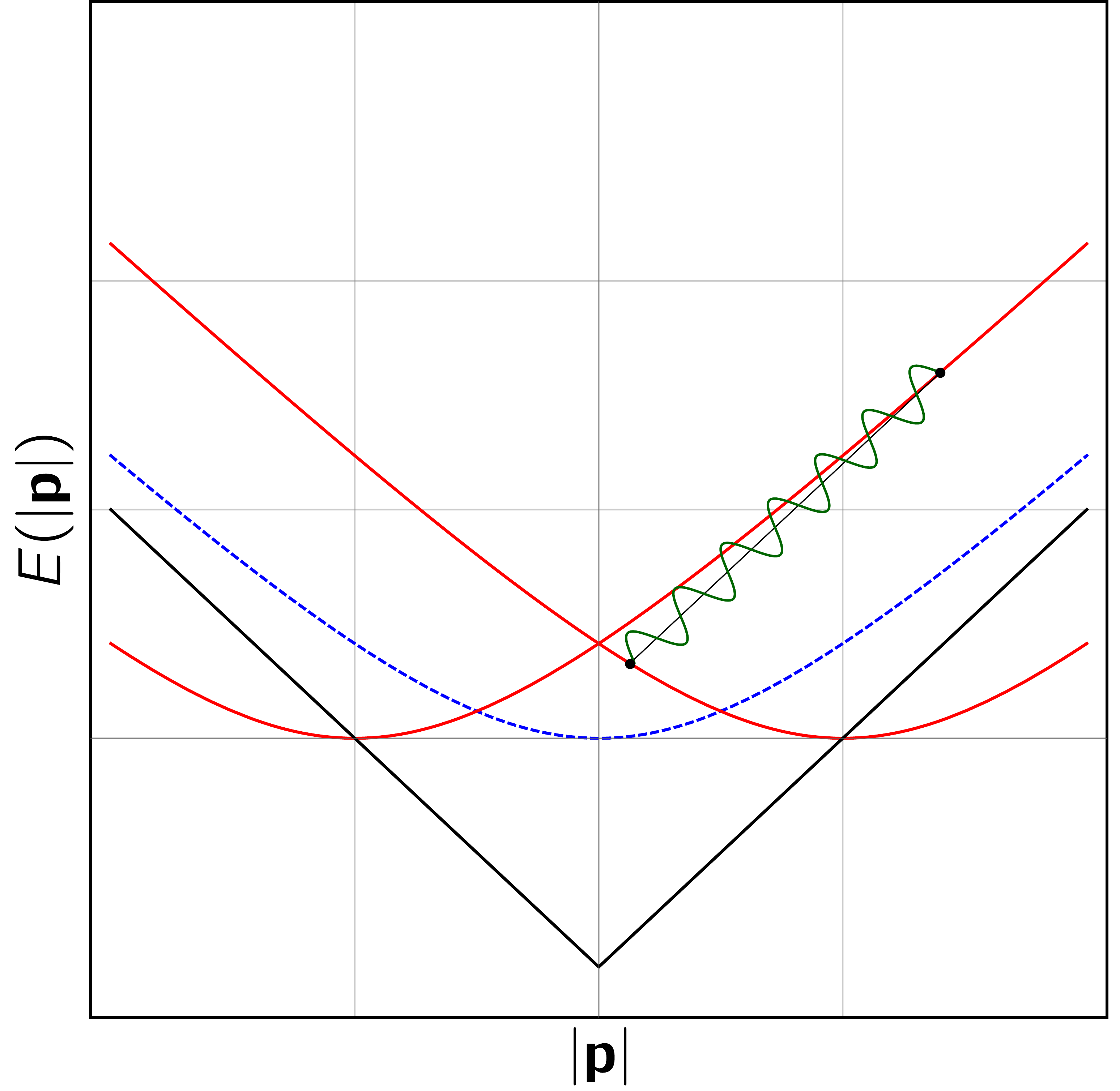}}
\caption{Standard nullcone (black), standard mass shell (blue, dashed), and modified mass shell (red) in momentum space. (\textbf{a}) isotropic spin-degenerate modification of Dirac theory with $c_{00}=-1/2$. From the initial energy of the modified fermion, two photon emissions are possible until the fermion stops radiating; (\textbf{b}) isotropic spin-nondegenerate modification of Dirac theory with $b_0=m_{\psi}$. There are two distinct modified dispersion laws. Photon emission can occur from one to the other branch.}
\label{fig:modified-fermions}
\end{figure}

To understand the physics of vacuum Cherenkov radiation qualitatively, it is reasonable to study the structure of the nullcone and mass shell in momentum space. This nullcone is a surface plot of the photon dispersion law $\omega=\omega(\mathbf{p})$ as a function of the momentum $\mathbf{p}$, whereas the mass shell is an analog surface plot of the massive dispersion law $E=E(\mathbf{p})$. For the studies that we are interested in, it is sufficient to consider one-dimensional slices of these surfaces as shown in Figures~\ref{fig:modified-photons} and \ref{fig:modified-fermions}.
Vacuum~Cherenkov radiation does not occur in the standard case simply due to the fact that the nullcone approaches the mass shell asymptotically for large momenta. Hence, it is not possible to shift one of the straight lines of the sliced nullcone such that it intersects the mass shell in two different~points.

Vacuum Cherenkov radiation can emerge either for Lorentz violation in the sector of force carriers or for Lorentz violation in the matter sector, i.e., when either the nullcone or the mass shell (or~both) are deformed. We will consider two examples of the first scenario. The first example is a quantum electrodynamics with an isotropic modification of the photon sector that is characterized by a controlling coefficient $\widetilde{\kappa}_{\mathrm{tr}}>0$. The modified nullcone structure is shown in Figure~\ref{fig:modified-photons}a. In this case, the~opening angle of the nullcone is larger in comparison to the standard theory, which means that photons travel with a phase velocity $v_{\mathrm{ph}}\equiv \omega(\mathbf{p})/|\mathbf{p}|<1$. Now, it is possible without any problems to draw lines parallel to the nullcone such that they intersect the mass shell at two distinct points. Therefore, the massive particle will lose energy by photon emission where the recoil due to the photon emitted can be large enough to reverse the direction of the particle momentum. An analog process with a final-state photon of lower energy can take place subsequently, as we can again construct a line parallel to the nullcone having two intersections with the mass shell. In the second emission process, the recoil of the emitted photon is not sufficient anymore to reverse the momentum direction of the massive particle. After the two photon emissions, the mass shell can no longer be intersected twice with lines parallel to the nullcone, i.e., the massive particle ceases to radiate photons. The second example in the scenario of modified photons is shown in Figure~\ref{fig:modified-photons}b. {It is based on an anisotropic Maxwell--Chern--Simons (MCS) term in the photon sector with dimensionful coefficient $k_{AF}^3=5m_{\psi},$ where $m_{\psi}$ is the fermion mass.} Here, the nullcone is totally deformed due to dispersion. Parts of the deformed nullcone can be shifted to connect pairs of points on the standard mass shell such that the lower points are identified with the minimum of the deformed nullcone. For both photon emissions {shown}, the recoil is large enough to reverse the momentum direction of the radiating particle. {The~construction illustrated in Figure~\ref{fig:modified-fermions}b can be perpetuated such that photons of ever increasing energy are emitted. This shows that vacuum Cherenkov radiation need not necessarily be a threshold~process.}

Vacuum Cherenkov radiation can also occur when the fermion sector is altered. The first example deals with a spin-degenerate modification of Dirac theory where the curvature of the modified mass shell is larger compared to the standard case, cf.~Figure~\ref{fig:modified-fermions}a. Two points on the mass shell can then be connected with lines parallel to the standard nullcone. Several photons may be emitted subsequently where the fermion momentum direction reverses by each emission. An alternative interesting possibility is that of a spin-nondegenerate modification of Dirac fermions shown in Figure~\ref{fig:modified-fermions}b. In such a case, the modified mass shell splits into two distinct branches dependent on the spin direction of the fermion with respect to the quantization axis. The process qualitatively differs from that illustrated in Figure~\ref{fig:modified-fermions}a, as one point sitting on a certain branch can be connected with a second point on the other branch by a line parallel to the standard nullcone. Photon emission takes place once and ceases when the second branch is reached. Furthermore, the direction of the fermion momentum does not reverse anymore, as a helicity change of the fermion is already caused by a spin flip due to the change of branch.

\section{Properties of Vacuum Cherenkov Radiation in Minkowski Spacetime}
\unskip
\subsection{Classical Description}

The presumably first detailed study of vacuum Cherenkov radiation was carried out by Coleman and Glashow in \cite{Coleman:1997xq}. Amongst other things, they consider an isotropic modification of the speed of light such that $c<1$. They find that massive particles start emitting photons once their energy lies above a certain threshold energy. The rate of energy loss $\mathrm{d}E/\mathrm{d}x$ is stated in the paper as well, and it goes to 0 when the particle energy approaches the threshold energy from above. Since they do not state a modified action, it is, first of all, not clear how they obtained $\mathrm{d}E/\mathrm{d}x$. A calculation of the energy loss requires knowledge of the dynamics of the process that can be inferred from a theory only. As will be explained later, the authors most probably considered the standard Feynman rules of a scalar quantum electrodynamics and a {phase space modified due to Lorentz violation}.

The first articles on vacuum Cherenkov radiation in the realm of the SME were written by Lehnert and Potting in 2004 \cite{Lehnert:2004be,Lehnert:2004hq}. They are based on a spacelike {MCS} electromagnetism coupled to standard massive particles. MCS theory is characterized by a dimensionless preferred direction $\zeta^{\mu}$ and a mass scale $m_{\mathrm{cs}}$. This framework is birefringent, i.e., it provides two distinct modified propagation modes~$\ominus$ and $\oplus$ of electromagnetic waves. Independently of the observer frame chosen, $\omega_{\ominus}<|\mathbf{k}|<\omega_{\oplus}$. The~properties of the dispersion laws are quite obscure in a purely timelike frame. For example, in~such a frame, the mode $\oplus$ has a cusp for a vanishing momentum and a gap exists where the energy of $\ominus$ becomes complex.

To simplify the analysis, the authors study the vacuum Cherenkov process in the rest frame of the massive particle, which is possible due to observer Lorentz invariance. In the rest frame, the four-current $j_{\mu}$ is stationary. The solution to the four-potential can be written as an integral over momentum space involving a product of the Green's function of MCS theory and the Fourier transform $\tilde{j}_{\mu}$ of the four-current:
\begin{equation}
A^{\mu}(\vec{r})=\int_{C_{\omega}}\frac{\mathrm{d}^3k}{(2\pi)^3}\,\frac{N^{\mu\nu}(\vec{k})\tilde{j}_{\nu}(\mathbf{k})\exp(\mathrm{i}\mathbf{k}\cdot\vec{r})}{D(0,\mathbf{k})}\,,
\end{equation}
where $C_{\omega}$ is an appropriate contour in the complex $|\mathbf{k}|$ plane and $N^{\mu\nu}$ is a tensor structure characteristic for MCS theory. The Green's function involves the inverse of the MCS dispersion equation $D(k^0,\mathbf{k})$ for $k^0=0$. The contour integral in the complex plane picks up contributions from the poles inside the contour only. If these poles are purely imaginary, the integrand will be exponentially damped for large distances from the trajectory. This means that the far-field vanishes whereupon there is no radiation far away from the source, which is the situation in standard electrodynamics. Hence, vacuum Cherenkov radiation occurs once there are real solutions of $D(0,\mathbf{k})=0$ (corresponding to spacelike dispersion relations) such that the exponential function is oscillatory. From the brief discussion of the modes of MCS theory above, we already know that such solutions exist indeed. They are given by the modes $\ominus$. The condition of real, spacelike solutions of the dispersion equation is equivalent to the existence of a refractive index $n>1$ of the vacuum. A boost to an arbitrary observer frame provides the condition on the Cherenkov angle: $\cos\theta_c=1/(n|\boldsymbol{\beta}|),$ where $\boldsymbol{\beta}$ is the three-velocity of the particle.

Furthermore, Lehnert and Potting found that the emitted radiation can be right-handed or left-handed polarized, dependent on the emission direction with respect to $\boldsymbol{\beta}$ and the spacelike preferred direction $\boldsymbol{\zeta}$. Using the energy-momentum tensor, it can be shown that there is no net radiated energy in the rest frame of a localized source. Furthermore, the radiated three-momentum vanishes due to the symmetry of the integrand as long as the integrand itself is regular. For real solutions of $D(0,\mathbf{k})=0$, the integrand is singular, in fact, and there are nonvanishing contributions to the radiated three-momentum. The latter has a backreaction on the radiating particle. Due to such recoil effects, the~particle will no longer move on a geodesic, i.e., the weak equivalence principle is violated under this condition.

Further results on vacuum Cherenkov radiation at the classical level were obtained by Altschul in a series of papers. Studies on the nonbirefringent part of the {\em CPT}-even modification of Maxwell theory were carried out in \cite{Altschul:2006zz}. These nine nonbirefringent coefficients are characterized by a symmetric, traceless matrix $\widetilde{\kappa}^{\mu\nu}$. Altschul finds that there exists a coordinate transformation between the $c^{\mu\nu}$ coefficients in the fermion sector and the $\widetilde{\kappa}^{\mu\nu}$ coefficients in the photon sector. Hence, investigating vacuum Cherenkov radiation in both sectors should produce equivalent results at first order in Lorentz violation. The possible problem of a missing frequency cutoff is identified in the computation of the total radiated energy. Therefore, integrating the radiated-energy spectrum $P(\omega)$ over the photon frequency $\omega$ may produce infinite results, as the modified theory allows for arbitrarily high $\omega$. A~hypothetical cutoff is introduced via the argument that there are possible causality problems of the theory in the vicinity of the threshold. However, we will see below that such a cutoff is obsolete when taking into account recoil effects. In the subsequent paper \cite{Altschul:2007kr}, Altschul studies additional properties of vacuum Cherenkov radiation for both MCS theory and modified Maxwell theory. He points out that MCS theory allows for a single subluminal mode only, whereas two such modes can arise in modified Maxwell theory dependent on the propagation direction. A consequence is that there is always a single Cherenkov cone only for MCS theory, whereas in modified Maxwell theory there may be two distinct cones for certain directions.

Timelike MCS theory is plagued by several problems. The energy density for the timelike sector is known to be unbounded from below. Thus, there exist runaway modes whose amplitudes increase exponentially as a function of the distance from the source. The possible origin of these issues is the gap of complex energies for the mode $\ominus$ indicated above. This gap exists for small momenta and long wavelengths, respectively, which is the motivation to study the power emitted by these long-wavelength modes in \cite{Altschul:2014bba}. The power of the short-wavelength modes is found by a phase space estimate to be proportional to $v^3$ at least (with the velocity $v$ of the radiating particle). As radiation of long wavelengths is likely to be emitted in the nonrelativistic regime and as these contributions cannot be estimated from the phasespace, they~must be of $\mathcal{O}(v^2)$. The timelike MCS theory leads to a modified Ampère law where the additional term $\mathbf{J}_{\mathrm{eff}}=2{\zeta^0}\mathbf{B}$ is interpreted as a current. {Altschul obtains the} perturbative solution of the modified Ampère law at leading order in {$\zeta^0$} and $v$ in a clever way. Based on this solution, the outgoing Poynting vector flux $\mathbf{S}\cdot\hat{\mathbf{e}}_r$ at $\mathcal{O}(v^2)$ is shown to be an odd function of $\cos\theta$ where $\theta$ is the polar angle. Hence, the total integral of $\mathbf{S}\cdot\hat{\mathbf{e}}_r$ over a closed surface vanishes demonstrating that the runaway modes do not contribute to vacuum Cherenkov radiation.

The latter result is generalized in \cite{Schober:2015rya} for the situation of a steady motion of the charge, i.e., recoil effects are neglected. The authors expand the electric and magnetic field $\mathbf{W}$ in terms of {$\zeta^0$} and the particle velocity $v$. They argue that each term of this expansion is either of toroidal nature, $W=W_r(r,\theta)\hat{\mathbf{e}}_r+W_{\theta}(r,\theta)\hat{\mathbf{e}}_{\theta}$, or of azimuthal form, $W=W_{\phi}(r,\theta)\hat{\mathbf{e}}_{\phi}$. The central finding is that the functions $W_r$, $W_{\theta}$, $W_{\phi}$ for the electric field $\mathbf{E}$, the magnetic field $\mathbf{B}$, and the vector potential $\mathbf{A}$ are either even or odd with respect to $\cos\theta$. As a result of that, each combination that contributes to the outgoing Poynting flux $\mathbf{S}\cdot\hat{\mathbf{e}}_r$ provides an odd function in $\cos\theta$. Therefore, any integral of $\mathbf{S}\cdot \hat{\mathbf{e}}_r$ over a sphere vanishes and there is no net outflow of energy at all!

{A possible interpretation of the latter result is that the negative energy carried away from the long-wavelength modes compensates the contributions of positive energy associated with the short-wavelength modes. Interestingly, a paper appeared very recently demonstrating this proposed cancelation explicitly \cite{DeCosta:2018nyf}. The authors work in the same framework as that developed in \cite{Schober:2015rya}. They~calculate the Fourier transform of the magnetic field at first order in $\zeta^0$ and $v$. Based on this finding, all magnetic field terms of odd orders in $\zeta^0$ can be obtained iteratively. Summing these contributions corresponds to a geometric series whose limit is an expression that is quadratic in the photon momentum and involves a singularity for the photon momentum equal to $2|\zeta^0|$. When the momentum approaches the singularity from above, there is a sign change of the whole expression demonstrating the cancelation between contributions of large momentum with those of low momentum. The singularity itself is not problematic, as its principal value can be considered formally.}

\subsection{Quantum Effects}

The presumably first studies of the vacuum Cherenkov process in the context of quantum field theory were carried out in \cite{Kaufhold:2005vj,Kaufhold:2007qd}. The first paper is based on a MCS theory coupled to standard Dirac fermions. Due to the issues of timelike MCS theory, an observer frame with a purely spacelike preferred direction $\boldsymbol{\zeta}$ is considered. The matrix element squared of the process at tree-level is obtained based on a set of modified Feynman rules. The decay rate follows from integrating the latter result over the modified phase space. As long as the fermion under consideration has a nonvanishing momentum component $q_{\parallel}$ parallel to $\boldsymbol{\zeta}$, the decay rate is found to be nonzero. For $|q_{\parallel}|\ll m_{\psi}$, the~decay rate depends on $q_{\parallel}$ in a polynomial manner, whereas for $|q_{\parallel}|\gg m_{\psi}$ it involves a logarithmic term. Such~logarithmic dependencies are nonperturbative in Lorentz violation and they seem to be characteristic for the ultra-high energy regime of a birefringent theory. An additional example for a behavior of this kind will be encountered below. Furthermore, the authors find that the emitted photon is circularly polarized for large photon momenta and linearly polarized for vanishing photon~momenta.

In their second paper \cite{Kaufhold:2007qd}, Klinkhamer and Kaufhold extend their analysis of \cite{Kaufhold:2005vj}. Apart from MCS theory coupled to standard Dirac fermions, they also consider a modified scalar quantum electrodynamics, i.e., they couple MCS theory to the Lagrange density of a massive, spinless field. Including the second framework into their investigation permits them to calculate the decay rates $\Gamma$ and radiated-energy rates $\mathrm{d}W/\mathrm{d}t$ for spin-1/2 fermions and spinless bosons, respectively. For fermion momenta $|q_{\parallel}|\ll m_{\psi}$, spin effects are found to be highly suppressed. However, at least for large fermion momenta, they play a certain role:
\begin{equation}
\label{eq:radiated-energy-rates}
\left.\frac{\mathrm{d}W_{\mathrm{scalar}}}{\mathrm{d}t}\right|_{|q_{\parallel}|\gg m_{\psi}}=\frac{\alpha}{4}m_{\text{cs}}|q_{\parallel}|+\dots\,,\quad \left.\frac{\mathrm{d}W_{\mathrm{spinor}}}{\mathrm{d}t}\right|_{|q_{\parallel}|\gg m_{\psi}}=\frac{\alpha}{3}m_{\text{cs}}|q_{\parallel}|+\dots\,,
\end{equation}
where $\alpha$ is the fine-structure constant.

It is interesting that these leading-order results can be obtained from a semiclassical description based on the following nontrivial refractive index of the vacuum:
\begin{equation}
\label{eq:refractive-index}
n(\omega)=1+\frac{m_{\text{cs}}|\cos\theta|}{2\omega}+\dots\,,
\end{equation}
with the photon energy $\omega$ and the angle $\theta$ between the photon momentum and $\boldsymbol{\zeta}$. Now, the photon spectrum for ordinary Cherenkov radiation in macroscopic media, which is provided by the Frank--Tamm formula, is reinterpreted in the context of a vacuum with the refractive index given by Eq.~(\ref{eq:refractive-index}).
Taking the photon momentum into account, which is a quantum effect, introduces a finite maximum energy of the emitted photon. Hence, the Frank--Tamm spectrum can be integrated over the photon frequency resulting in the first radiated-energy rate of Eq.~(\ref{eq:radiated-energy-rates}). This finding makes sense, as Frank and Tamm did not consider the {spin of the radiating particle}. The latter leads to an extra contribution that adds up to the second rate of Eq.~(\ref{eq:radiated-energy-rates}).

A further step towards a better understanding of vacuum Cherenkov radiation in the quantum regime is achieved via \cite{Klinkhamer:2008ky}. In the latter reference, isotropic modified Maxwell theory based on the single coefficient $\widetilde{\kappa}_{\mathrm{tr}}$ is coupled to standard Dirac fermions. The process is found to have a threshold $E_{\mathrm{th}}\approx m_{\psi}/\sqrt{2\widetilde{\kappa}_{\mathrm{tr}}}$. Furthermore, the decay {rate and radiated-energy rate} are computed again {using} a set of modified Feynman rules for photons and a modified phase space. For comparison, a semiclassical analysis is performed in addition based on a modified refractive index $n(\omega)=1+\widetilde{\kappa}_{\mathrm{tr}}+\dots$ of the vacuum. Recoil effects lead to a finite radiated-energy rate that is found to be equal to the corresponding result in \cite{Coleman:1997xq} for large momenta. The deduction from this finding is that Coleman and Glashow {also} did not take into account the particle spin. By doing so, the leading-order classical result is equal to the leading-order rate of the quantum field theory analysis including Dirac fermions.

{At around the same time, the paper \cite{Hohensee:2008xz} was published. The latter also deals with isotropic modifications of both the photon and the fermion sector. In a quantum field theory analysis, the~authors derive both the threshold and the decay rate for the vacuum Cherenkov process at leading order in Lorentz violation. These results are consistent with those of \cite{Klinkhamer:2008ky} at leading order. A constraint on isotropic Lorentz violation in the photon and electron sector follows from the assumption that an energy loss by vacuum Cherenkov radiation hides within the uncertainty of the synchrotron losses at the Large Electron-Positron Collider (LEP).}

A couple of years later, D\'{i}az and Klinkhamer refined {the study carried out in \cite{Diaz:2015hxa}.} Here,~the~ parton model is used to understand effects related to a possible internal structure of the incoming fermion. In this approach, a Cherenkov photon can be emitted from a single parton inside a fermion, i.e.,~neutrons~can radiate vacuum Cherenkov radiation, as well. The authors determine the emission rate by an electrically charged parton carrying the fraction $xp$ of the incoming fermion momentum $p$. The~maximum photon energy is found to be proportional to the fraction $x$. Hence, Cherenkov emission is suppressed for small momentum fractions, which practically excludes this process for the {sea quarks}. Another interesting property is that emissions of very hard photons may lead to momentum transfers large enough to destroy the initial hadron and to produce new ones in the final state. Finally, the total power radiated by a composite proton and neutron, respectively, is determined based on very recent data on parton distribution functions. The rates for both an initial proton and neutron are suppressed by around one order of magnitude in comparison to the rate in \cite{Klinkhamer:2008ky} for a structureless Dirac fermion. This result does not change the constraint obtained in \cite{Klinkhamer:2008ky} though.

A quantum field theory analysis that complements the classical studies of \cite{Altschul:2014bba,Schober:2015rya,DeCosta:2018nyf} for vacuum Cherenkov radiation in timelike MCS theory is carried out in \cite{Colladay:2016rmy}. Complex energies of the mode~$\ominus$ can be avoided by introducing a nonvanishing photon mass. Note that the current experimental limit on a photon mass is $m_{\upgamma}<\unit[1\times 10^{-27}]{GeV}$ {\cite{Tanabashi:2018}}, which is many orders of magnitudes larger than the best constraint on the controlling coefficient under consideration, which is $\zeta^0\lesssim \unit[10^{-43}]{GeV}$ {\cite{Kostelecky:2008ts}}. A~nonzero photon mass also allows for a consistent quantization procedure whereupon the decay rate for vacuum Cherenkov radiation can be calculated in quantum field theory. The process has a threshold and the decay rate for initial fermion momenta much larger than this threshold is characterized by two different regimes. There is a range of momenta where it is suppressed quadratically by Lorentz violation and increases quadratically with the fermion momentum. For very large momenta, it is only linearly suppressed by Lorentz violation and grows linearly with momentum.

After the big amount of investigations of vacuum Cherenkov radiation with modified photons, time was ready to study the vacuum Cherenkov process for Lorentz-violating fermions in \cite{Schreck:2017isa}. The~coordinate transformation moving Lorentz violation between the photon and fermion sector~\cite{Altschul:2006zz} only covers the {\em CPT}-even, nonbirefringent $c$ coefficients {for fermions}. Such a transformation is neither known for the {\em CPT}-odd $a$, $b$, $e$, $f$, and~$g$ coefficients nor for the {\em CPT}-even, birefringent $d$, $H$ coefficients, which is why such an analysis is definitely reasonable. Note also that for the birefringent $b$, $d$, $H$, and~$g$ coefficients, there are two fermion modes that will be denoted as $\oplus_X$ and $\ominus_X$ where the index $X$ stands for a controlling coefficient. This property {leads} to interesting new effects that do not have a counterpart in the photon sector. The base of the analysis performed in \cite{Schreck:2017isa} {are} the modified Feynman rules for external fermion legs developed in the technical work \cite{Reis:2016hzu}. Due to calculational complexities, decay rates and radiated-energy rates are obtained numerically for a broad range of coefficients. The essential properties are that the processes for the $c$, $d$, $e$, $f$, and $g$ coefficients have thresholds. Furthermore, for large incoming fermion momenta, the decay rates for the $c$, $d$ coefficients are linearly suppressed in Lorentz violation, whereas the decay rates for $e$, $f$, and $g$ are quadratically suppressed. The processes $\oplus_{d,g}\rightarrow \oplus_{d,g}+\upgamma$ are allowed for a certain sign of the controlling coefficients and $\ominus_{d,g}\rightarrow \ominus_{d,g}+\upgamma$ occur for the opposite sign. Analog decays for the $a$, $b$, and $H$ coefficients are found to be forbidden. However, the birefringent nature of the spin-nondegenerate coefficients opens up the possibility of spin--flip processes $\oplus_{b,d,H,g}\rightarrow \ominus_{b,d,H,g}+\upgamma$ and the analog ones with $\oplus$ and $\ominus$ interchanged. These decays have peculiar properties. First, they do not have a threshold and, second, their decay rates are highly suppressed by several powers of the controlling coefficients in contrast to the spin-conserving decays. Furthermore, the decay rates involve logarithmic terms for large momenta similarly to the results for MCS theory obtained in \cite{Kaufhold:2005vj}.

\subsection{Radiation of Particles Other Than Photons}

A plethora of alternative Cherenkov-type processes is conceivable in Lorentz-violating theories. The radiating particle does not necessarily have to be a massive fermion and the radiated particles need not inevitably be photons. One of the first articles in this {realm} was published by Altschul \cite{Altschul:2016ces}. Here, an effective theory for Lorentz-violating pions is considered coupled to standard photons. In~this framework, high-energy photons can lose energy by radiating pions where the process involves a threshold. At leading order in Lorentz violation, the threshold has parallels with the analog expression for nonbirefringent modified Maxwell theory in \cite{Altschul:2006zz}. After all, both frameworks are quite similar from a merely kinematic point of view. As the direct calculation of the decay rate is cumbersome, an~estimate of the rate is obtained from the prominent process $\Gamma(\uppi^0\rightarrow 2\upgamma)$.

A further interesting Cherenkov-type process is explored in \cite{Colladay:2017qfr}. The authors consider a Lorentz-violating modification of the free W-boson sector. Such a term enables a Cherenkov-type process in vacuo where a fermion of mass $m_1<m_{\mathrm{W}}$ decays into a fermion of mass $m_2<m_{\mathrm{W}}$ under the emission of a W boson. An analysis in the rest frame of the incoming particle is feasible. It reveals that the process is energetically possible for a quite large controlling coefficient in the rest frame showing that this frame must be strongly boosted with respect to a concordant frame where Lorentz violation is supposed to be small. Analogous to \cite{Diaz:2015hxa}, the authors perform a sophisticated parton model study based on the elementary process mentioned before. Hence, the decay rate of W-boson emission of a single quark is convoluted with appropriate parton distribution functions. As the probability of finding quarks with a large momentum fraction $xp$ steeply decreases for $x\mapsto 1$, the decay rate of a proton right at the threshold region is considerably suppressed in comparison to the decay rate of the elementary process.

After the detection of superluminal neutrinos {had} been announced by the OPERA collaboration in 2011, Cohen and Glashow quickly published a paper studying Cherenkov-type radiation of electron-positron pairs by neutrinos via virtual Z bosons \cite{Cohen:2011hx}. Based on the value for Lorentz violation in the neutrino sector announced by OPERA, they showed that neutrinos would quickly lose their energy via this Cherenkov process making it impossible for them to arrive at the Gran Sasso laboratory with the energy that was measured by OPERA. {Few months later, OPERA announced that errors had occurred in the course of the measurement process rendering their results incorrect.}

\section{Cherenkov-Type Radiation in Modified Gravity}

In this review, special emphasis shall be put on Cherenkov-type processes in theories of modified gravity. Before doing so, it is reasonable to recall certain properties of General Relativity that are crucial in the context of Lorentz violation. We will refer to the Einstein equations without the cosmological constant that are given by
\begin{subequations}
\begin{align}
G_{\mu\nu}&=8\pi G_N(T_M)_{\mu\nu}\,, \\[2ex]
G_{\mu\nu}&={R_{\mu\nu}-\frac{R}{2}g_{\mu\nu}\,.}
\end{align}
\end{subequations}

Here, $G_{\mu\nu}$ is the Einstein tensor that decomposes into the Ricci tensor $R_{\mu\nu}$ and the Ricci scalar $R$. The spacetime metric is given by $g_{\mu\nu}$ and $(T_M)_{\mu\nu}$ stands for the energy-momentum tensor of matter. The coupling constant of gravity and matter corresponds to Newton's constant $G_N$.

\subsection{Lorentz Violation in Gravity}

General Relativity exhibits invariance under diffeomorphisms and under general coordinate transformations where these concepts are equivalent. A diffeomorphism is a differentiable map of a manifold onto itself whose inverse has the same properties. Diffeomorphism invariance can also be understood as a gauge symmetry of the theory. It is possible to choose a gauge fixing condition appropriately to remove four of the ten independent components of the spacetime metric. Hence,~only~six~metric components may possibly be physical. Since one set of Bianchi identities of Riemannian geometry provides $D_{\mu}G^{\mu\nu}=0$ with the covariant derivative $D_{\mu}$, an alternative argument is that the latter four identities eliminate four spacetime metric components. Thus, we see that the Bianchi identities are closely connected to diffeomorphism invariance. Diffeomorphisms in Minkowski spacetime simply correspond to translations. After all, translations map Minkowski spacetime onto~itself.

The global Lorentz invariance of Special Relativity is lost in General Relativity. However, there still exists the concept of local Lorentz invariance. To understand what that means, it is useful to formulate General Relativity with the vierbein {approach}. The vierbein (tetrad) allows for transforming between the coordinate basis of the spacetime manifold and the basis of a local inertial (freely falling) reference frame at each spacetime point. The great advantage of dealing with the vierbein {formalism} is that it allows us to introduce Lorentz invariance as a local concept, i.e., a symmetry that is valid in a local frame \cite{Kostelecky:2003fs}.

In the context of gravity, matter in curved spacetimes is described by minimally coupling appropriate fields to the Einstein--Hilbert action. This means that the Minkowski metric has to be replaced by a curved spacetime metric wherever it occurs. Furthermore, ordinary partial derivatives are replaced by covariant derivatives on the manifold. Dependent on whether the metric or vierbein formalism is used, the determinant of the metric and the vierbein, respectively, has to be taken into account in the volume element of the action. Furthermore, the vierbein formalism is indispensable when introducing Dirac fermions into the theory due to their additional spinor structure. As it is clear how to deal with the spinor structure in a local inertial frame, the vierbein formalism extends the concept of a spinor from such a local frame to the frame of spacetime coordinates.
\begin{figure}[t]
\centering
\subfloat[]{\label{fig:original-sphere}\includegraphics[scale=0.6]{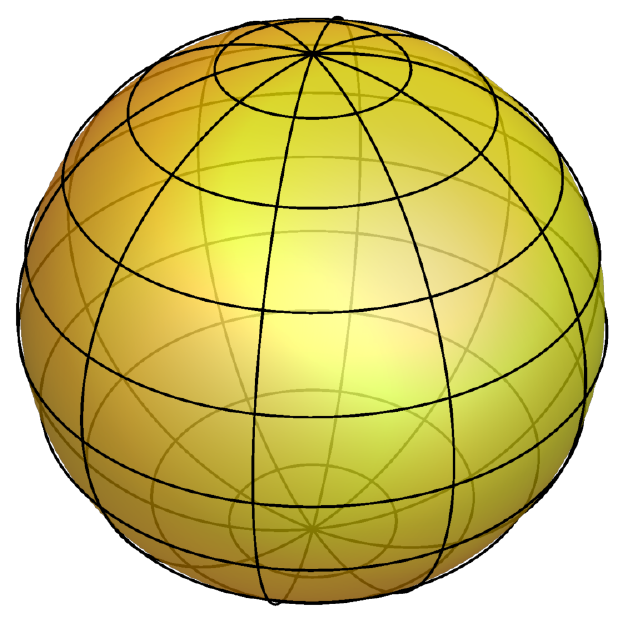}}\hspace{1cm} \subfloat[]{\label{fig:general-coordinate-transformation}\includegraphics[scale=0.6]{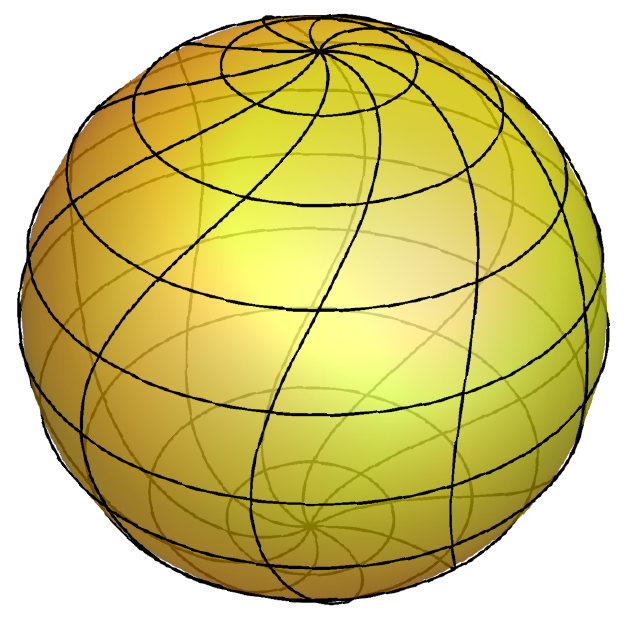}}\hspace{1cm}
\subfloat[]{\label{fig:diffeomorphism}\includegraphics[scale=0.6]{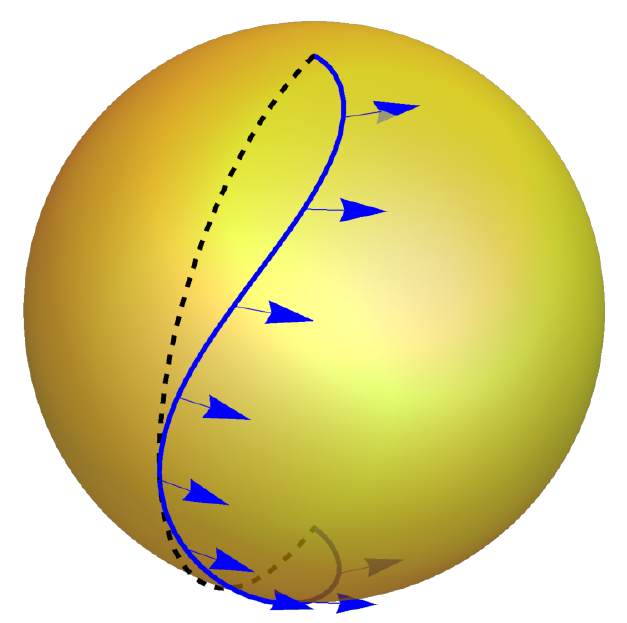}} \\
\subfloat[]{\label{fig:original-sphere-with-vectorfield}\includegraphics[scale=0.6]{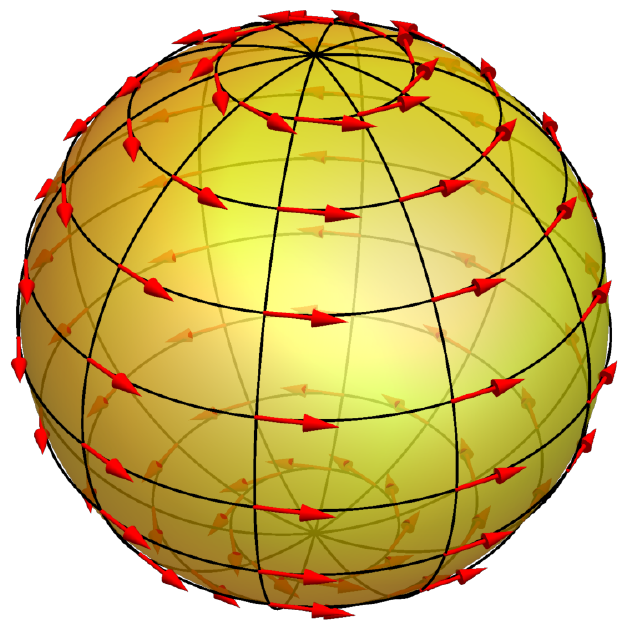}}\hspace{1cm}
\subfloat[]{\label{fig:general-coordinate-transformation-vector-field}\includegraphics[scale=0.6]{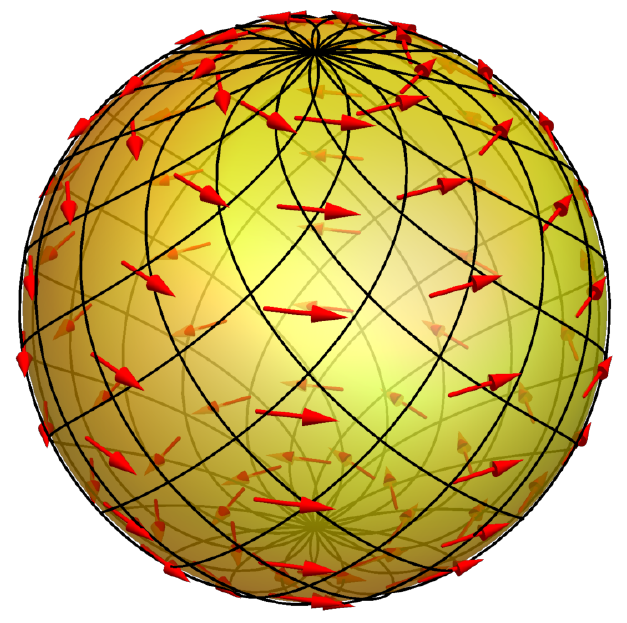}}\hspace{1cm}
\subfloat[]{\label{fig:diffeomorphism-vector-field}\includegraphics[scale=0.6]{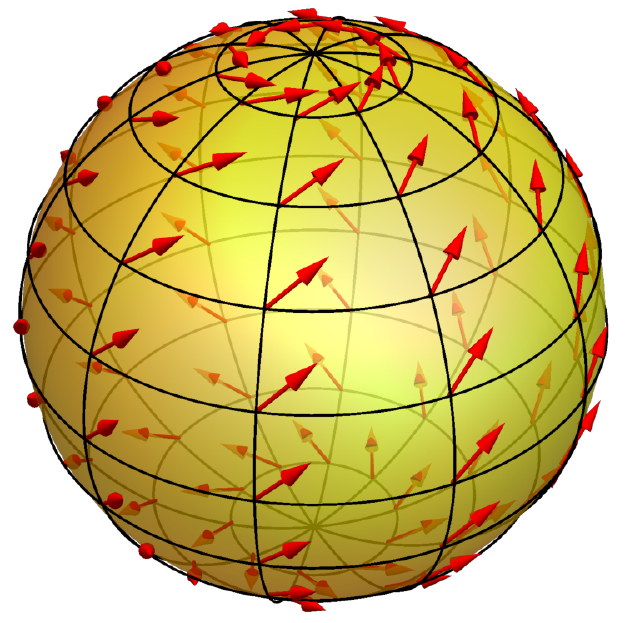}}
\caption{(\textbf{a}) two-sphere ($S^2$) parameterized in terms of the polar angle $\vartheta$ and the azimuthal angle $\varphi$. The coordinate lines of constant $\vartheta$ are circles parallel to the equator where the coordinate lines of constant $\varphi$ are great circles linking the north and south poles with each other; (\textbf{b}) illustration of a general coordinate transformation for $S^2$ that deformes the original coordinate lines; (\textbf{c}) illustration of a diffeomorphism for $S^2$. To distinguish it from a general coordinate transformation, the unchanged coordinate lines are not shown. The graphics demonstrates how an original geodesic connecting the north and the south pole gets deformed by the diffeomorphism; (\textbf{d}) $S^2$ endowed with an additional vector field that could correspond to a matter field in physics; (\textbf{e})~general coordinate transformation applied to $S^2$. The coordinate lines are deformed, but~the vector field itself stays untouched. As the coordinates change, however, the explicit representation of the vector field changes as well; (\textbf{f}) diffeomorphism applied to $S^2$ with an intrinsic vector field. In contrast to before, the coordinates remain unchanged, but the vector field transforms in a nontrivial manner.}
\label{fig:coordinate-transformation-diffeomorphism}
\end{figure}

\begin{figure}
\centering
\subfloat[]{\label{fig:original-plane}\includegraphics[scale=0.3]{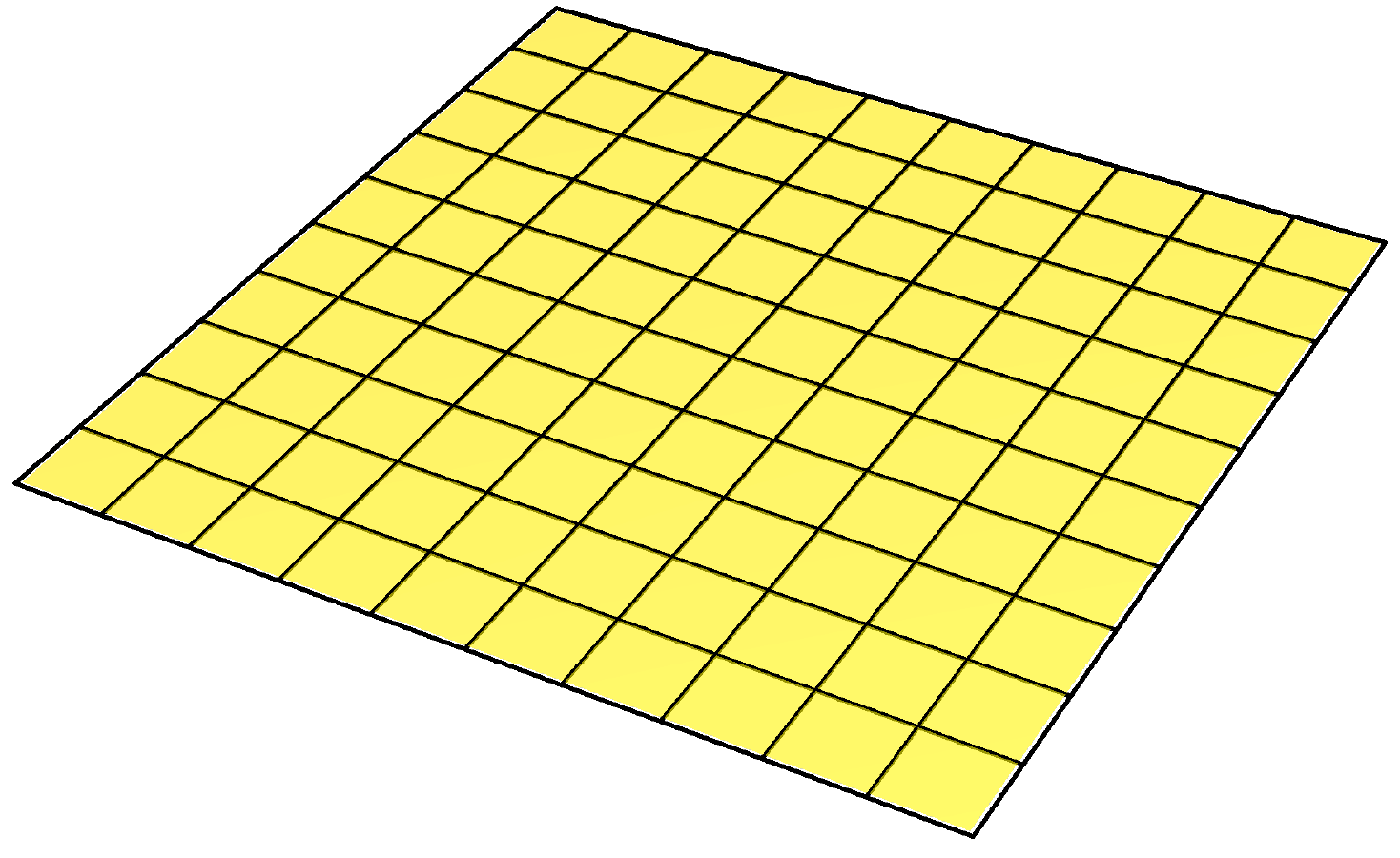}}\hspace{1cm} \subfloat[]{\label{fig:plane-coordinate-transformation}\includegraphics[scale=0.3]{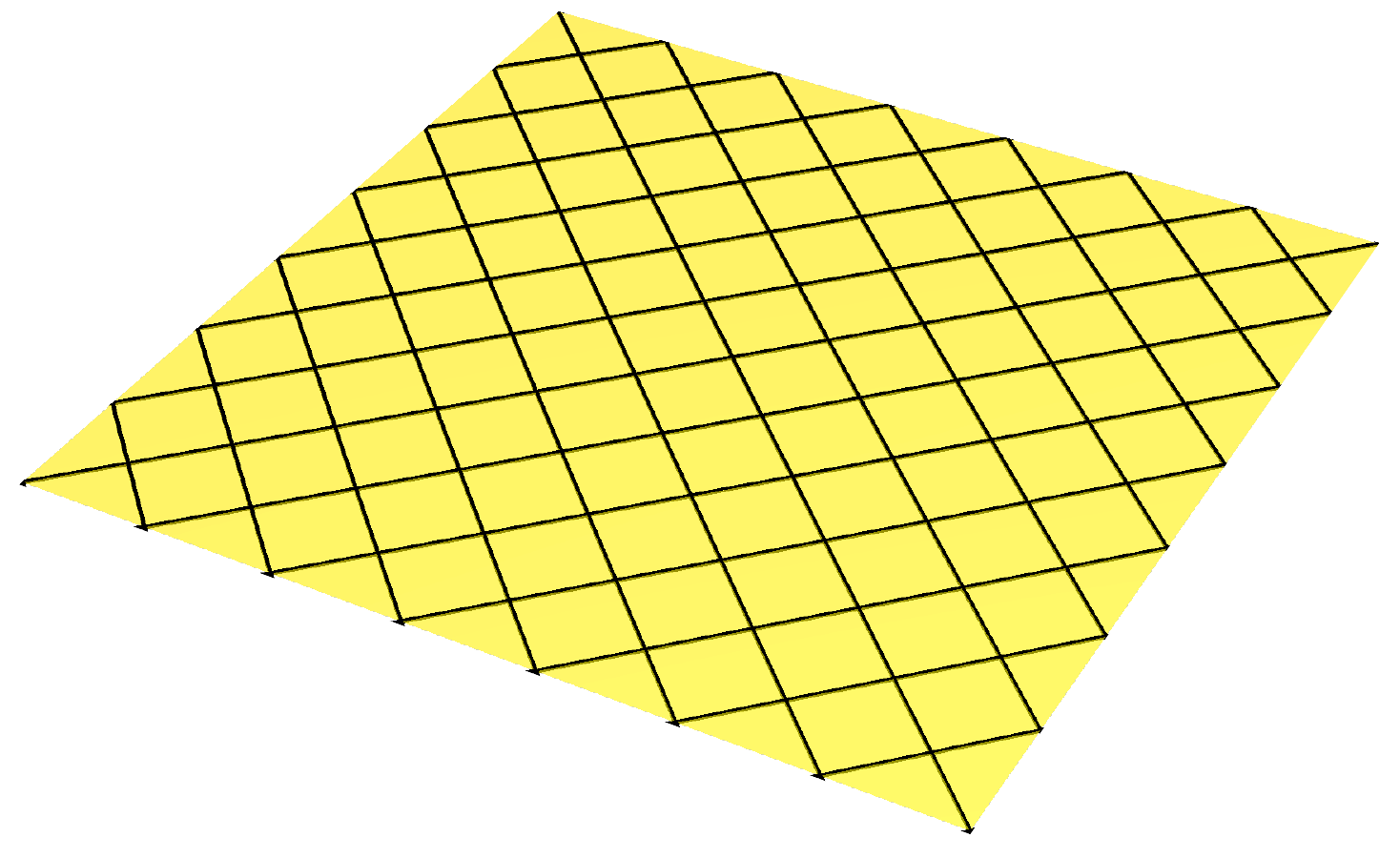}}\hspace{1cm}
\subfloat[]{\label{fig:plane-diffeomorphism}\includegraphics[scale=0.3]{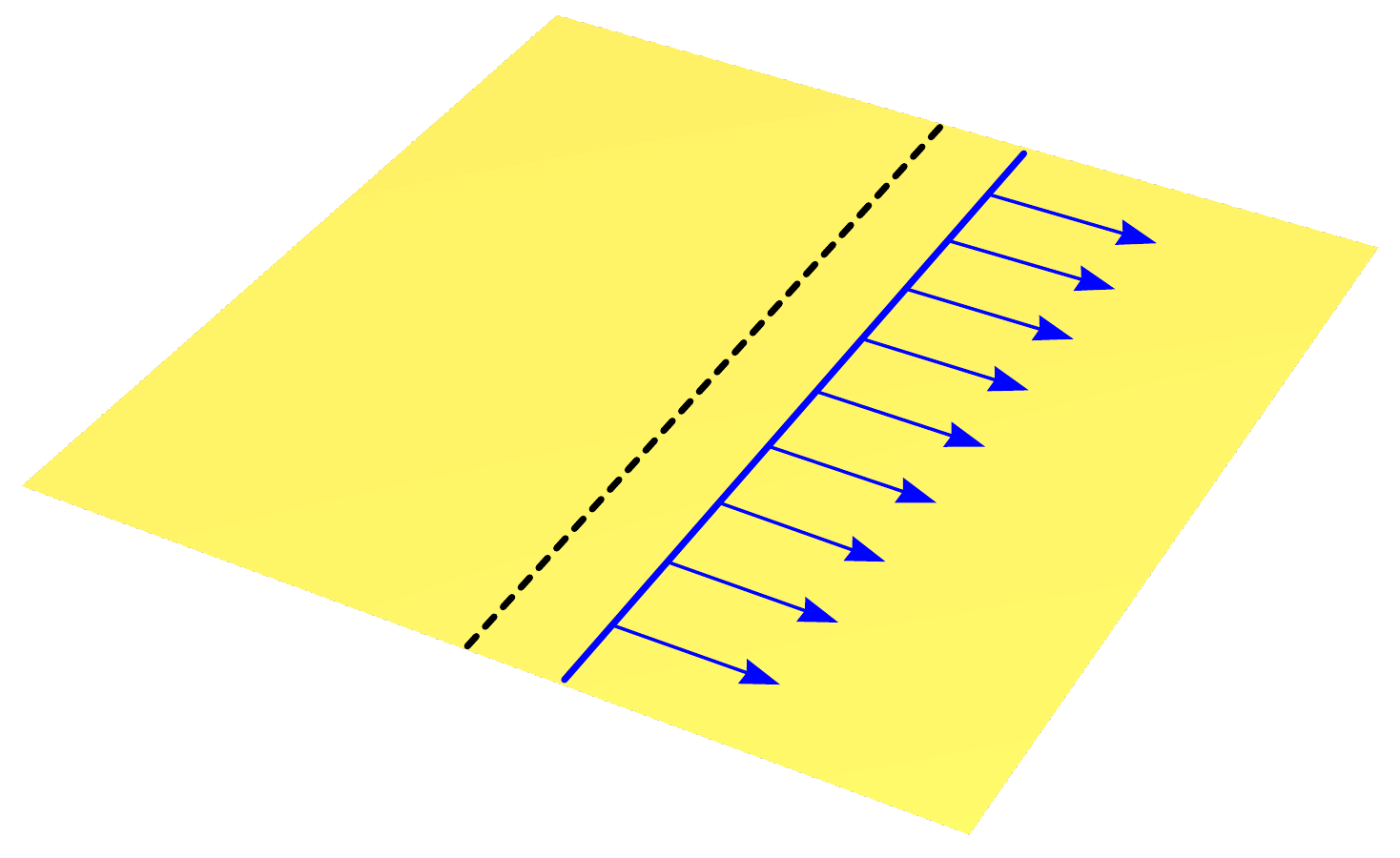}} \\
\subfloat[]{\label{fig:plane-vectorfield}\includegraphics[scale=0.3]{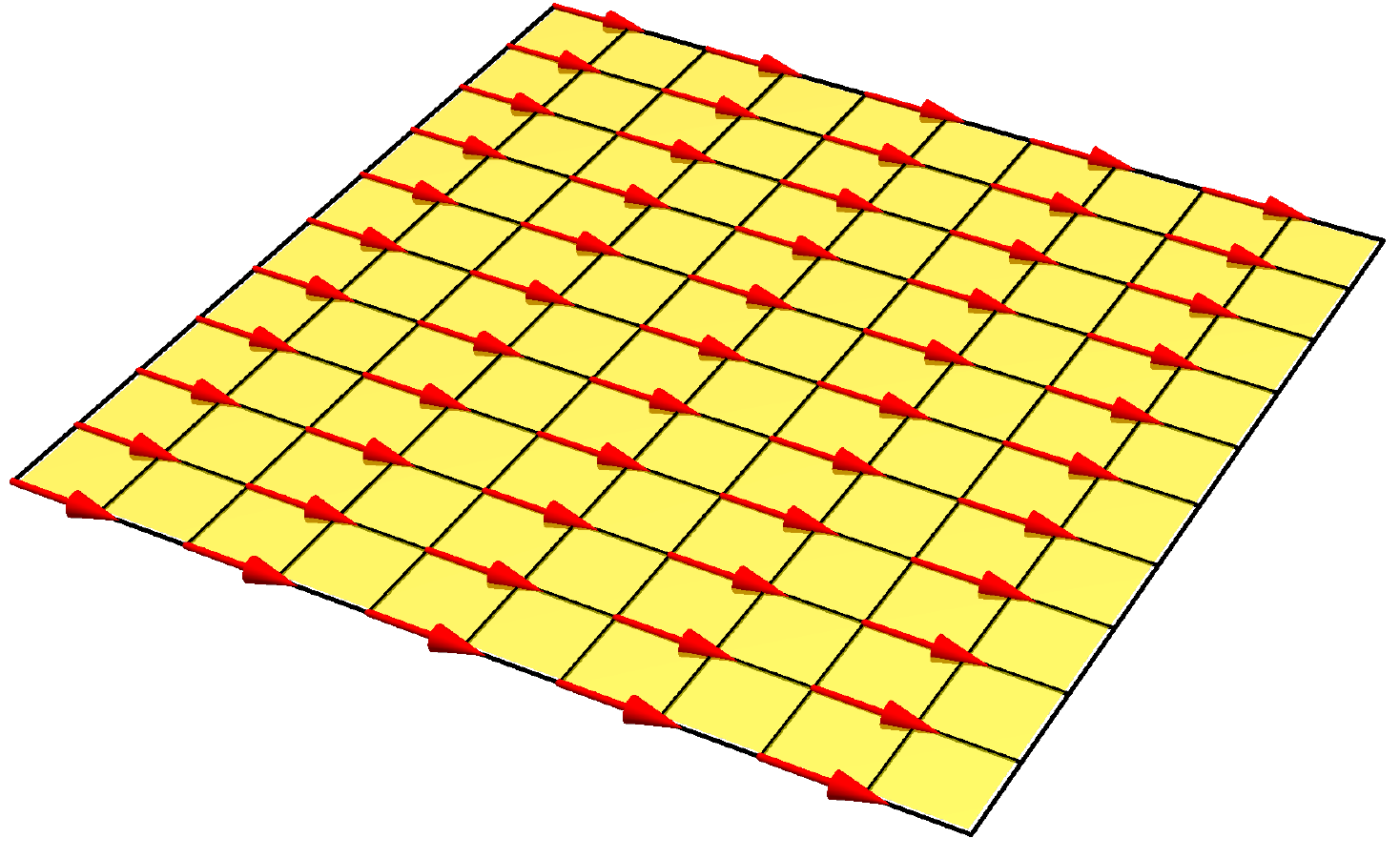}}\hspace{1cm}
\subfloat[]{\label{fig:plane-observer-lorentz-transformation}\includegraphics[scale=0.3]{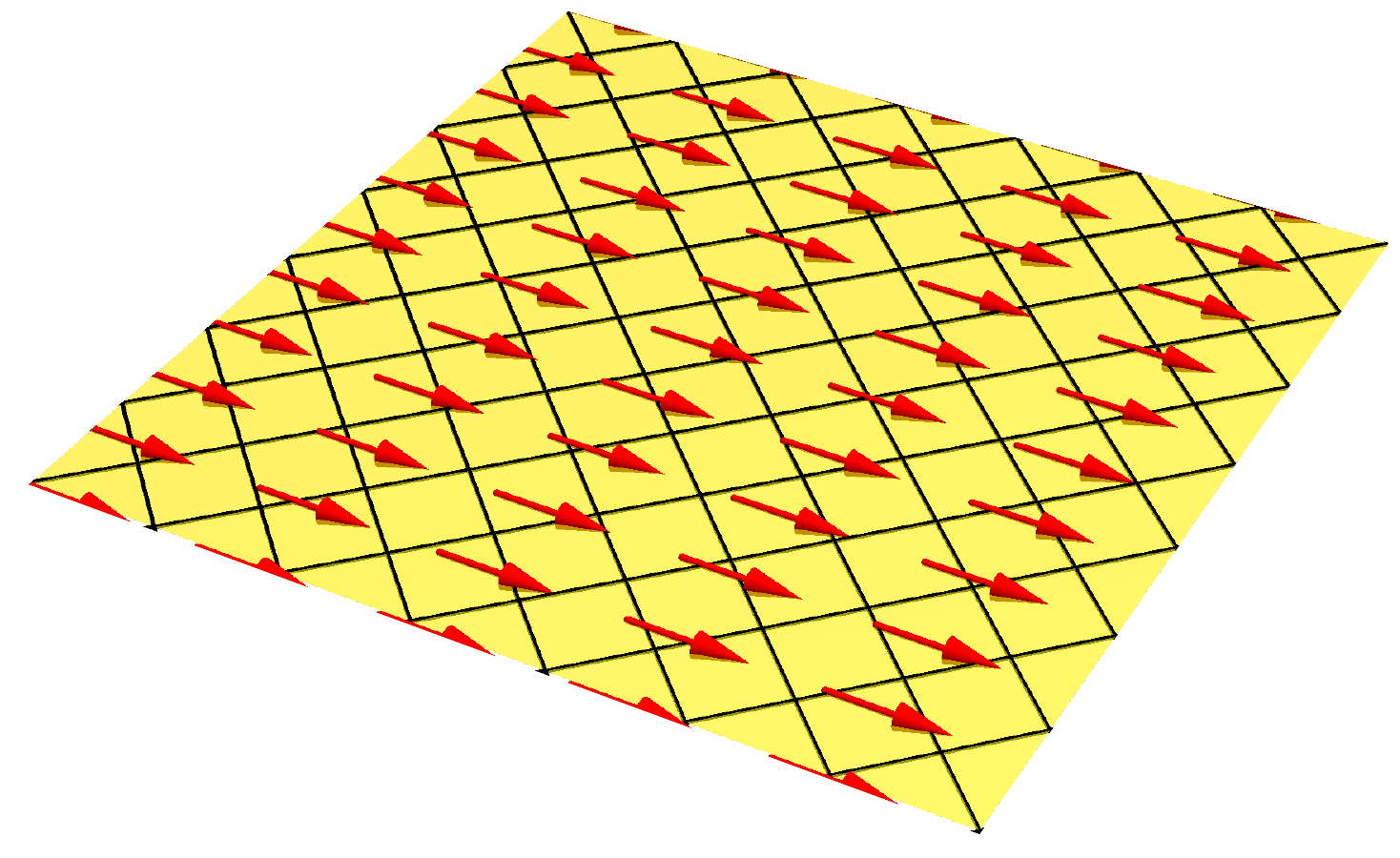}}\hspace{1cm}
\subfloat[]{\label{fig:plane-particle-lorentz-transformation}\includegraphics[scale=0.3]{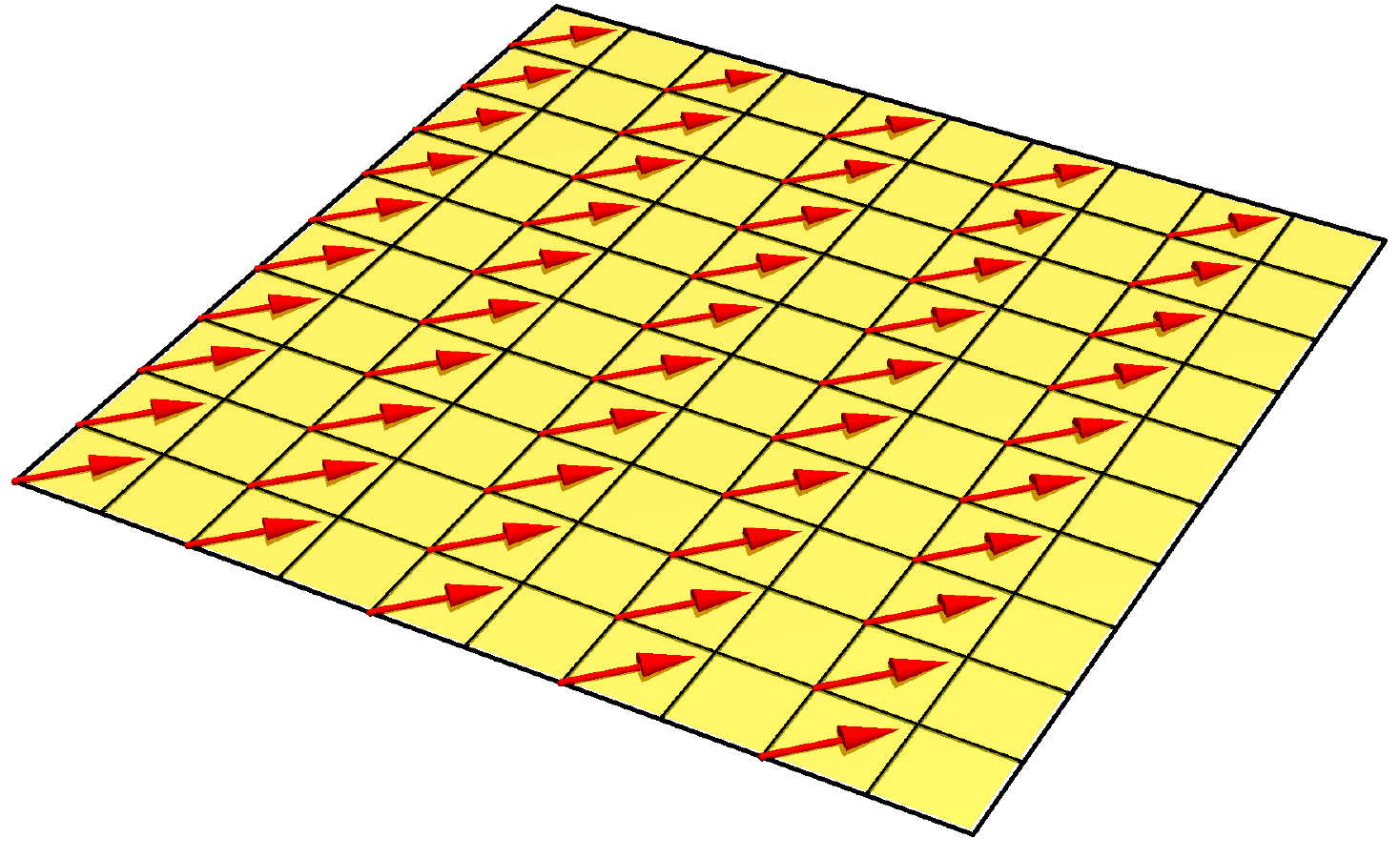}}
\caption{(\textbf{a}) tangent plane of $S^2$ with coordinate lines shown; (\textbf{b}) observer Lorentz transformation rotating the coordinate lines; (\textbf{c}) diffeomorphism (translation) that simply maps the points of a geodesic (straight line) to the points of a parallel geodesic. {The unmodified coordinate lines are omitted for clarity;} (\textbf{d}) tangent plane endowed with a vector field; (\textbf{e}) observer Lorentz transformation that changes the coordinate lines, but leaves the vector field untouched; (\textbf{f}) particle Lorentz transformation that induces a nontrivial transformation of the vector field. The coordinate lines are not modified.}
\label{fig:coordinate-transformation-diffeomorphism-tangent-plane}
\end{figure}

There are two types of local Lorentz transformations: observer and particle transformations. With matter coupled to General Relativity, it is easier to clarify their implications. Observer Lorentz transformations correspond to rotations and boosts of a local reference frame, whereas particle Lorentz transformations deal with rotations and boosts of matter fields in such local frames. (Note that the SME community avoids the terms passive and active transformations that are actually used in the literature. One of the reasons is that an active transformation may be supposed to transform both the background field and the matter fields leaving the coordinates unchanged. In contrast, a particle transformation touches the matter field only, whereas the background field remains fixed.)
The term general coordinate transformations already suggests that this type of transformations only changes the choice of coordinates of the base manifold. On the contrary, diffeomorphisms transform the matter fields including the metric tensor, which in the context of General Relativity is considered as a field as well. Therefore, general coordinate transformations as well as observer Lorentz transformations merely change the description of the physics, but leave the physics itself untouched. On the other hand, both diffeomorphisms and particle Lorentz transformations act on the physical fields though. Diffeomorphisms and general coordinate transformations are equivalent in General Relativity and so are particle and observer Lorentz transformations. However, when including Lorentz violation into the theory, these concepts have to be distinguished from each other carefully, cf.~Figures~\ref{fig:coordinate-transformation-diffeomorphism} and \ref{fig:coordinate-transformation-diffeomorphism-tangent-plane} for an illustration on a curved manifold and its tangent plane, respectively.

The gravitational part of the SME is developed in \cite{Kostelecky:2003fs} and its formulation is based on the vierbein formalism. In the gravitational SME, the conceptual differences between diffeomorphisms and general coordinate transformations as well as particle and observer Lorentz transformations become evident. The SME introduces background fields that break local particle Lorentz invariance, i.e., there may be one or several preferred directions in each freely falling reference frame. The background field does not transform like a tensor anymore under particle Lorentz transformations, but it remains fixed. However, there is still the freedom to choose coordinates within these frames as desired. A violation of local particle Lorentz invariance leads to a violation of diffeomorphism invariance and vice versa. The reason is that any coordinate-dependent vector field on a curved manifold has a homogeneous counterpart in a local inertial reference frame at each point of the manifold and vice versa. As local observer Lorentz invariance is still granted, so is invariance under general coordinate transformations.

In the gravitational SME, the minimal matter-sector SME contributions, which were originally formulated in Minkowski spacetime, are minimally coupled to a curved spacetime. {The vierbein must be accompanied by} an object known as the spin connection granting that the covariant derivative of the vierbein vanishes. The spin connection and the affine connection (Christoffel symbols) are not independent from each other. Apart from Lorentz violation in the matter sector, the~gravitational SME describes Lorentz violation in the pure-gravity sector. The~dimension-4 contributions incorporate a single Lorentz-invariant term multiplied by the curvature scalar, a~symmetric matrix of nine~nonbirefringent coefficients contracted with the Ricci tensor, and~10~birefringent coefficients contracted with the Weyl tensor, which is the totally traceless Riemann tensor~\cite{Kostelecky:2003fs}.

{In addition to} the properties outlined above, there is a further highly subtle statement on the origin of Lorentz violation in gravity. First of all, the Lorentz-violating background fields in Minkowski spacetime are usually taken as independent from the spacetime coordinates, which implies energy-momentum conservation. This choice is reasonable from a physical perspective, since a constant background can be considered as the dominant contribution and a part dependent on the spacetime coordinates is interpreted as a correction. In the context of a curved spacetime, however, it is not clear what ``constant'' means. A first justified thought may be to require that the covariant derivative of a background field vanish. However, there are only few manifolds that permit such a requirement and those are not that interesting in the context of gravity. Anyhow, background fields in gravity will have some dependence on the spacetime coordinates. Therefore, Lorentz violation in the matter sector leads to modifications of the conservation law of the energy-momentum tensor where the latter now includes nonvanishing spacetime derivatives of the controlling coefficients. However, the identity $D_{\mu}G^{\mu\nu}=0$ is still valid even in the presence of Lorentz violation, as it simply follows from the underlying Riemannian framework. Hence, there is a clash when including Lorentz violation into a gravity theory by hand (explicit Lorentz violation) \cite{Kostelecky:2003fs}. The way out of this no-go theorem is to consider spontaneous Lorentz violation, i.e., Lorentz-violating background fields that have a dynamical origin and satisfy field equations on their own. Therefore, the action of the gravitational SME must also include terms that describe possible mechanisms providing a way of how to break local Lorentz invariance spontaneously.

An explicit violation of Lorentz invariance in a theory of modified gravity is analogous to an explicit violation of gauge invariance in a gauge theory. The latter is obviously considered as destructive for the fate of a gauge theory because it leads to a breakdown of unitarity, the production of unphysical modes, etc. In contrast to that, a spontaneous violation of gauge invariance is a tremendously useful physical mechanism capable of explaining a multitude of effects both in particle physics and in condensed-matter physics. For a spontaneously violated gauge symmetry, the underlying theory is still perfectly gauge-invariant. It is just the ground state of the theory that randomly picks a certain gauge configuration, cf.~the vacuum expectation value of the Higgs field in the Higgs mechanism. In the context of Lorentz violation, a background field emerges as vacuum expectation values of a tensor-valued fundamental field. Hence, the underlying fundamental theory is Lorentz invariant, but~the ground state picks certain preferred directions in spacetime. As the background field is dynamic, picking a definite ground state is accompanied by fluctuations, i.e., both Nambu--Goldstone modes and massive modes are created. The first correspond to small fluctuations of the preferred directions around the direction chosen by the vacuum expectation values. The second are associated with fluctuations of the magnitude of controlling coefficients.

\subsection{Gravitational Vacuum Cherenkov Radiation}

After reviewing some of the crucial implications of Lorentz violation in gravity, we are now ready to discuss the characteristics of the Cherenkov-type process in gravity that is investigated in \cite{Kostelecky:2015dpa}. In~their paper, Kosteleck\'{y} and Tasson consider the nonbirefringent coefficients of the pure-gravity sector. In contrast to \cite{Kostelecky:2003fs}, all operators of even mass dimension $d\geq 4$ are taken into account. Therefore, this analysis is currently the only one including contributions from operators of higher mass dimension. For the purpose of studying Cherenkov-type processes in gravity, it suffices to restrict their analysis to linearized gravity, i.e., the metric perturbation $h_{\mu\nu}$ is neglected at second and higher orders in the field equations \cite{Kostelecky:2016kfm,Kostelecky:2017zob}. The metric perturbation has 10 components out of which four are gauge degrees of freedom, as discussed before. Another four components are auxiliary and two are physical ones that propagate. The latter two correspond to the polarization modes of a gravitational wave.

Lorentz violation is restricted to first order as well. The corresponding coefficients are assumed to be constant and small. In principle, they are put in by hand, which is why Lorentz invariance breaking is explicit in this treatment. However, the coefficients considered are chosen such that gauge symmetry (diffeomorphism invariance) is restored at linear order \cite{Kostelecky:2016kfm}. (In \cite{Kostelecky:2017zob}, the claim for gauge invariance is relaxed and sets of coefficients are included that violate gauge invariance.)
Therefore, the number of polarization modes of a gravitational wave remains fixed and merely Lorentz-violating perturbations of the standard polarizations are expected. Both the gauge and auxiliary components are removed by appropriately adapted gauge fixing conditions. These are modifications of the standard Hilbert gauge and transverse traceless gauge conditions that are usually employed in the context of linearized~gravity.

The resulting Einstein equations have the form of a wave equation for the metric perturbation $h_{\mu\nu}$. The solutions of these equations are interpreted as gravitational waves perturbed by Lorentz violation. A description of this kind introduces a nontrivial refractive index of the vacuum. In the current context, the latter is a refractive index for gravitational waves instead of electromagnetic waves. At linear order in Lorentz violation, the refractive index involves a sum over all possible Lorentz-violating contributions of even mass dimension $d\geq 4$ where the coefficients for $d>4$ are contracted with additional wave four-vectors. Therefore, the modified gravity theory exhibits anisotropy, but no dispersion for $d=4$, whereas, for $d>4$, both anisotropy and dispersion play a role.

Within a semiclassical approach, a gravitational wave is quantized in the sense that it carries a momentum in analogy to a photon resulting from the quantization of electromagnetic waves. Quanta of the linearized gravitational field are introduced as well, and they are known as gravitons. Within this semiclassical treatment of linearized gravity, Kosteleck\'{y} and Tasson derive modified perturbative Feynman rules for distinct particle sectors. They consider graviton emission from spinless bosons, photons, and Dirac fermions. Each of these vertices is proportional to the square root of Newton's constant $G_N$, i.e., the matrix element squared is directly proportional to $G_N$ and quadratically suppressed by Lorentz violation, in addition.

The notion of conserved quantities along geodesics is a very subtle one in General Relativity. It requires some knowledge of isometries of the underlying curved spacetime metric encoded in the Killing vectors. However, local energy-momentum conservation directly follows from Einstein's equations combined with $\partial_{\mu}G^{\mu\nu}=0$. Since interactions occur locally in the language of perturbation theory and Feynman rules, energy-momentum conservation at each vertex is granted. Hence, kinematic arguments {such as} those outlined in Section~\ref{sec:fundamentals-kinematics} also apply.

The radiated-energy rate follows from integrating the matrix element squared over an appropriately modified phase space. Due to {the nontrivial} refractive index {of the vacuum}, particles~may travel faster than the phase velocity of gravity in vacuo, rendering graviton emission possible. This~process is {of Cherenkov-type and} can occur for all known particles, as they all interact with gravity. Analogous to conventional vacuum Cherenkov radiation, a Cherenkov angle can be introduced as the angle between the propagation direction of the radiating particle and the gravitons radiated. The~Cherenkov angle is defined only above a certain velocity of the incoming particle demonstrating that gravitational Cherenkov radiation is a threshold process. Another interesting property is that energy-momentum conservation introduces an upper cutoff for the energy of the emitted gravitons. The latter corresponds approximately to the incoming particle momentum. Thus, the phase space integral is rendered finite without any issues. Recall the analog behavior found for vacuum Cherenkov radiation in Minkowski spacetime \cite{Kaufhold:2005vj,Kaufhold:2007qd,Klinkhamer:2008ky}.

In the ultrarelativistic limit, integration over phase space for any of the three processes considered simply delivers a factor $F^w(d)$ that depends on the mass dimension of the Lorentz-violating operators. Here, $w=\{\phi,\psi,\gamma\}$ for scalars, Dirac fermions, and photons, respectively. In general, the energy loss per time has the form
\begin{equation}
\label{eq:energy-loss-rate}
\frac{\mathrm{d}E}{\mathrm{d}t}=-F^w(d)G_N(s^{(d)})^2|\mathbf{p}|^{2d-4}\,,
\end{equation}
with the incoming particle momentum $\mathbf{p}$ and the Lorentz-violating operator $s^{(d)}$, which is a direction-dependent combination of coefficients of mass dimension $d$. The factors $F^w(d)$ have the property that $F^{\phi}<F^{\psi}<F^{\gamma}$, where $F^{\phi}\sim 16/d$, $F^{\psi}\sim 8$, and  $F^{\gamma}\sim 16d$ for $d\mapsto\infty$. Hence, a photon loses more energy by gravitational Cherenkov radiation than a Dirac fermion and a Dirac fermion more than a spinless particle. The reason for this behavior is to be found in the spin of the radiating particle. A graviton is a spin-2 excitation, i.e., its spin projection $S_{\xi}$ with respect to some quantization axis $\hat{\boldsymbol{\xi}}$ can take each value out of the set $S_{\xi}=\{-2,-1,0,1,2\}$. Hence, the emission of gravitons with $S_{\xi}=\pm 2$ is only possible for photons of polarizations $\lambda_+=1$ and $\lambda_-=-1$, respectively. A complete flip of the polarization $\lambda_{\pm}\mapsto\lambda_{\mp}$ during the process can deliver $S_{\xi}=\pm 2$ necessary to produce such gravitons. For Dirac fermions, only gravitons with $S_{\xi}=\{-1,0,1\}$ can be emitted where for scalars $S_{\xi}=0$ is the remaining possibility. Since there are less decay channels for Dirac fermions in comparison to photons, energy loss of fermions is suppressed. A similar argument holds for scalars in comparison to fermions.

The main objective of \cite{Kostelecky:2015dpa} is to place stringent bounds on Lorentz violation in the pure-gravity sector. From the behavior of $F^w(d)$, we see that high-energy cosmic photons have the largest potential of doing so. However, as cosmic fermions with much higher energy are observed, the analysis in \cite{Kostelecky:2015dpa} is based on fermions. To simplify it, the authors consider operators of a particular mass dimension at a time. To constrain isotropic Lorentz violation, a single cosmic-ray event is {sufficient}. However, to carry out the same for a whole set of anisotropic coefficients, several such events coming from different directions are needed. It is convenient to state these directions in terms of azimuthal and polar angles on the celestial sphere. Hence, it also makes sense to expand the controlling coefficients in spherical harmonics. A {considerable} compilation of high-energy cosmic-ray events {is} chosen. To obtain conservative constraints, Kosteleck\'{y} and Tasson base their analysis on iron nuclei where a graviton is assumed to be emitted by one of the fermionic partons. In contrast to \cite{Diaz:2015hxa}, they do not consider parton distribution functions explicitly. Instead, the fermionic parton emitting gravitons is taken to have 10\% of the total incoming cosmic-ray energy. A set of bounds for a given $d$ is obtained by a maximizing procedure to fulfill the requirement that Lorentz violation can only be as large as to not obstruct cosmic rays with a certain energy from arriving on Earth.

By doing so, it is possible to constrain Lorentz violation for $d=4$, 6, and 8. The isotropic constraints are one-sided, whereas the anisotropic ones are two-sided. The rough order of magnitude of the constraints is $10^{-13}$ for $d=4$, $\unit[10^{-29}]{GeV^{-2}}$ for $d=6$, and $\unit[10^{-45}]{GeV^{-4}}$ for $d=8$. These bounds are impressive and are currently the best ones in the pure-gravity sector. {As constraints on Lorentz violation in the neutrino sector are the tightest ones, they can serve as a reference for comparisons. Very recently, the ICECUBE collaboration published a paper including novel neutrino bounds obtained from the observation of ultra-high energy neutrinos \cite{Aartsen:2017ibm}. The constraints for the dimension-6 neutrino coefficients are better than those in the gravity sector by several orders of magnitude, whereas the sensitivities for the dimension-8 coefficients can compete with each other.} If ultrahigh-energy protons can clearly be identified as primaries of cosmic-ray air showers, {the gravity sector} bounds have a tremendous potential of improvement.

\section{Conclusions}

The intention of the current review was to provide a summary of the crucial concepts of vacuum Cherenkov radiation developed over the past 20 years. At this point, it makes sense to take a look at the set of constraints on SME coefficients obtained from the absence of vacuum Cherenkov radiation, which~are compiled in Table~\ref{fig:compilation-cherenkov-constraints}. Most of the bounds are on dimensionless coefficients {except for} the constraints in the W boson and the graviton sector partially. A subset of the bounds in the graviton sector are the only ones on nonminimal coefficients showing that most studies on vacuum Cherenkov radiation so far have been restricted to the minimal SME. The birefringent coefficients of both MCS theory and modified Maxwell theory have never been constrained via the absence of vacuum Cherenkov radiation. The reason is that this type of coefficients can be much more tightly bounded by the absence of vacuum birefringence.

The isotropic $c$ coefficients and the isotropic $\widetilde{\kappa}_{\mathrm{tr}}$ are linked by the coordinate transformation between the photon and the matter sector found in \cite{Altschul:2006zz}. Furthermore, the $f$ coefficients squared correspond to $c$ coefficients \cite{Altschul:2006ts}. Hence, bounds are usually placed on combinations of these coefficients. The related ones in the proton sector are quite stringent, whereas the bound in the electron sector is weaker. As free electrons are not expected as a component of high-energy cosmic rays, the latter constraint {is based on the maximum energy reached at LEP} \cite{Hohensee:2008xz}. The constraints for the $e$ and $g$ coefficients are also weaker, since the related threshold energies depend inversely on these coefficients instead of via an inverse square root of the coefficient {\cite{Schreck:2017isa}}.

As the bound on Lorentz violation in the minimal gravity sector is dimensionless, it can be compared to the remaining dimensionless constraints. Although the results obtained in the pure-gravity sector are currently the best ones, they are weaker than the best dimensionless constraints in the photon and fermion sector. The reason is that gravity involves the dimensionful coupling constant $G_N$ in comparison to the other three fundamental interactions. Hence, the energy loss of Eq.~(\ref{eq:energy-loss-rate}) is directly proportional to $G_N$ and needs two additional powers of the incoming fermion momentum to cancel the inverse mass dimensions of $G_N$. The additional powers of the momentum are not enough to compensate for the smallness of $G_N$. Furthermore, Eq.~(\ref{eq:energy-loss-rate}) is not just linear in Lorentz violation, but~depends quadratically on it.

The future perspective in this interesting subfield is to continue studying Cherenkov-type processes in vacuo based on Lorentz violation of the nonminimal SME. With the detection of cosmic-ray events of higher energy and an improved identification of primaries in air showers, there is a great potential {of even better tests of} the very foundations of both the Standard Model and General Relativity.
\begin{table}[H]
\centering
\setlength\extrarowheight{1pt}
\begin{tabular}{cccc}
\toprule
\textbf{Sector} & \textbf{Constraint}	& \textbf{Reference} \\
\midrule
{Proton} & $-5\times 10^{-23}<\ring{c}^{\mathrm{UR}(4)}$ & \cite{Coleman:1997xq} \\[1pt]
{Proton, Photon} & $\widetilde{\kappa}_{\mathrm{tr}}-(4/3)c_{00}^{(4)\mathrm{p}}<6\times 10^{-20}$ & \cite{Klinkhamer:2008ky} \\[1pt]
{Electron, Photon} & $\widetilde{\kappa}_{\mathrm{tr}}-(4/3)c_{00}^{(4)\mathrm{e}}<1.2\times 10^{-11}$ & \cite{Hohensee:2008xz} \\[1pt]
{Gravity} & {$\overline{s}_{00}^{(4)}>-3\times 10^{-14}$} & \cite{Kostelecky:2015dpa} \\[1pt]
{Gravity} & {$|\overline{s}_{jm}^{(4)}|\lesssim 10^{-13}$} & \cite{Kostelecky:2015dpa} \\[1pt]
{Gravity} & {$\overline{s}_{00}^{(6)}<\unit[2\times 10^{-31}]{GeV^{-2}}$} & \cite{Kostelecky:2015dpa} \\[1pt]
{Gravity} & {$|\overline{s}_{jm}^{(6)}|\lesssim \unit[10^{-29}]{GeV^{-2}}$} & \cite{Kostelecky:2015dpa} \\[1pt]
{Gravity} & {$\overline{s}_{00}^{(8)}>\unit[-7\times 10^{-49}]{GeV^{-4}}$} & \cite{Kostelecky:2015dpa} \\[1pt]
{Gravity} & {$|\overline{s}_{jm}^{(8)}|\lesssim \unit[10^{-45}]{GeV^{-4}}$} & \cite{Kostelecky:2015dpa} \\[1pt]
{Pion} & $\delta^{\pi}>-7\times 10^{-13}$ & \cite{Altschul:2016ces} \\[1pt]
{Quark, Photon} & $-3\times 10^{-23}\leq c_{00}^{(4)\mathrm{u}}-(3/4)\widetilde{\kappa}_{\mathrm{tr}}-(3/8)(f_0^{(4)\mathrm{u}})^2$ & \cite{Schreck:2017isa} \\[1pt]
{Quark} & $|d_{00}^{(4)\mathrm{u}}|<3\times 10^{-23}$ & \cite{Schreck:2017isa} \\[1pt]
{Quark} & $|e_0^{(4)\mathrm{u}}|<9\times 10^{-12}$ & \cite{Schreck:2017isa} \\[1pt]
{Quark} & $|\ring{g}_1^{(4)\mathrm{u}}|<9\times 10^{-12}$ & \cite{Schreck:2017isa} \\[1pt]
{W boson} & $|(k_1)^{\mu}|<\unit[1.7\times 10^{-8}]{GeV}$ & \cite{Colladay:2017qfr} \\[1pt]
{W boson} & $|(k_2)^{\mu}|<\unit[1.1\times 10^{-7}]{GeV}$ & \cite{Colladay:2017qfr} \\
\bottomrule
\end{tabular}
\caption{Compilation of constraints on SME coefficients obtained from the absence of Cherenkov-type radiation in the vacuum.
Most of the constraints can be found in the data tables \cite{Kostelecky:2008ts}. {As the number of constraints in the gravity sector is extensive, approximate constraints on groups of coefficients are given where $0<j\leq d-2$ (with the mass dimension $d$) and $0\leq m\leq j$.}}
\label{fig:compilation-cherenkov-constraints}
\end{table}


\funding{This research was funded by CNPq grant number 421566/2016-7 and by FAPEMA grant number 01149/17.}

\acknowledgments{It is a pleasure to thank the Brazilian agencies CNPq and FAPEMA for their financial support. Furthermore, the author is indebted to the two anonymous referees
for useful comments on the submitted version of the paper.}

\conflictsofinterest{The author declares no conflict of interest.}
\reftitle{References}





\end{document}